\documentclass[10pt,twocolumn]{article}
\usepackage[a4paper,margin=2cm]{geometry}
\usepackage{authblk}
\usepackage{abstract}
\usepackage{titlesec}

\usepackage{abstract}
\usepackage{titlesec}

\usepackage{amsthm}

\makeatletter
\renewcommand{\@maketitle}{%
  \begingroup
  \renewcommand{\thefootnote}{\fnsymbol{footnote}}%
  \begin{center}%
    {\large\bfseries \@title \par}%
    \vskip 0.7em%
    {\normalsize
      \lineskip .4em%
      \begin{tabular}[t]{c}%
        \@author
      \end{tabular}\par}%
    \vspace{-0.2em}
    {\small (Dated: \@date) \par}%
  \end{center}%
  \vskip 0.5em%
  \@thanks
  \endgroup
}
\makeatother

\renewcommand{\thesection}{\Roman{section}}

\renewcommand{\thesubsection}{\Alph{subsection}}

\titleformat{\section}
  {\centering\normalfont\bfseries}
  {\thesection.}
  {1em}
  {\MakeUppercase}

\titleformat{\subsection}
  {\centering\small\bfseries}
  {\thesubsection.}
  {1em}
  {}

\titleformat{\subsubsection}
  {\normalfont\itshape}
  {\thesubsubsection.}
  {0.5em}
  {}

\titlespacing*{\section}
  {0pt}
  {4.0ex plus 1.0ex minus 0.3ex}
  {2.0ex plus 0.4ex}

\titlespacing*{\subsection}
  {0pt}
  {4.0ex plus 1.0ex minus 0.3ex}
  {2.0ex plus 0.4ex}

\titlespacing*{\subsubsection}
  {0pt}
  {1.2ex plus 0.4ex minus 0.2ex}
  {0.8ex plus 0.2ex}

\usepackage[utf8]{inputenc}
\usepackage{enumerate}
\usepackage{braket}
\usepackage{amsmath,dsfont}

\usepackage{amssymb}
\usepackage{svg}
\usepackage{upgreek}
\usepackage{amsfonts}

\usepackage{ragged2e}

\usepackage{graphicx}
\usepackage{quantikz}
\usepackage{tikz}
\usepackage{xcolor}

\definecolor{stimRed}{HTML}{f67a7b}
\definecolor{stimBlue} {HTML}{89bcf9}

\newcommand{\opname}[1]{%
  \wire["\scalebox{2}{\textcolor{red}{$#1$}}"{above,pos=1}]{a}%
}
\newcommand{\highlightop}[1]{%
  \wire[
    "\colorbox{yellow!40}{\scalebox{2}{\textcolor{red}{$#1$}}}"
    {above,pos=1}
  ]{a}%
}
\newcommand{\boxmark}[1]{%
  \gategroup[
    17,
    steps=#1,
    style={
      dashed,
      rounded corners,
      fill=blue!10,
      inner xsep=-1pt
    },
    background
  ]{}%
}
\newcommand{\largetext}[1]{\scalebox{2}{$#1$}}
\usepackage[caption=false]{subfig}

\makeatletter
\long\def\@makecaption#1#2{%
  \par
  \vskip\abovecaptionskip
  \begingroup
    \small
    \justifying
    #1. #2\par
  \endgroup
  \vskip\belowcaptionskip
}
\makeatother
\usepackage{graphicx}
\usepackage{pgffor}
\usepackage{comment}
\theoremstyle{plain}
 
\usepackage{float}

\usepackage{fancyhdr}
\usepackage{xcite}
\usepackage{color}
\usepackage{xcolor}
\usepackage{pifont}

\usepackage{MnSymbol}
\usepackage{bm}
\usepackage{bigints}

\usepackage{placeins}

\newcommand{\cbox}[1]{\colorbox{#1}{\rule{1ex}{1ex}}}

\usepackage[
  style=numeric-comp,
  sorting=none,
  sortcites=true,
  backend=biber,
  bibencoding=utf8,
  maxbibnames=3,
  minbibnames=3,
  giveninits=true,
  doi=false,
  isbn=false,
  url=true
]{biblatex}

\usepackage{hyperref}
\usepackage[capitalize]{cleveref}

\DeclareNameAlias{author}{family-given}

\DefineBibliographyStrings{english}{
  andothers = {et\addabbrvspace al\adddot}
}

\DeclareFieldFormat{title}{#1}
\DeclareFieldFormat[article]{title}{#1}
\DeclareFieldFormat[book]{title}{#1}
\DeclareFieldFormat[inproceedings]{title}{#1}

\DeclareFieldFormat[article]{volume}{\mkbibbold{#1}}
\DeclareFieldFormat[book]{volume}{\mkbibbold{#1}}
\DeclareFieldFormat[inproceedings]{volume}{\mkbibbold{#1}}

\DeclareFieldFormat{pages}{#1}

\DeclareFieldFormat{url}{}
\DeclareFieldFormat{doi}{}

\renewbibmacro*{in:}{}

\newbibmacro*{venue-volume-pages-year}{%
  \iffieldundef{journaltitle}
    {%
      \iffieldundef{booktitle}
        {%
          \iflistundef{publisher}
            {}
            {\printlist{publisher}}%
        }
        {\printfield{booktitle}}%
    }
    {\printfield{journaltitle}}%
  \iffieldundef{volume}
    {}
    {%
      \setunit{\addspace}%
      \printfield{volume}%
    }%
  \iffieldundef{pages}
    {}
    {%
      \setunit{\addcomma\space}%
      \printfield{pages}%
    }%
  \setunit{\addspace}%
  \printtext[parens]{\printfield{year}}%
}

\newbibmacro*{linked-venue-volume-pages-year}{%
  \iffieldundef{url}
    {%
      \iffieldundef{doi}
        {%
          \usebibmacro{venue-volume-pages-year}%
        }
        {%
          \href{https://doi.org/\thefield{doi}}{%
            \usebibmacro{venue-volume-pages-year}%
          }%
        }%
    }
    {%
      \href{\thefield{url}}{%
        \usebibmacro{venue-volume-pages-year}%
      }%
    }%
}

\DeclareBibliographyDriver{article}{%
  \printnames{author}%
  \setunit{\addcomma\space}%
  \printfield{title}%
  \setunit{\addcomma\space}%
  \usebibmacro{linked-venue-volume-pages-year}%
  \finentry
}

\DeclareBibliographyDriver{book}{%
  \printnames{author}%
  \setunit{\addcomma\space}%
  \printfield{title}%
  \setunit{\addcomma\space}%
  \usebibmacro{linked-venue-volume-pages-year}%
  \finentry
}

\DeclareBibliographyDriver{inproceedings}{%
  \printnames{author}%
  \setunit{\addcomma\space}%
  \printfield{title}%
  \setunit{\addcomma\space}%
  \usebibmacro{linked-venue-volume-pages-year}%
  \finentry
}
\DeclareRobustCommand{\emailstar}{%
  \textsuperscript{\protect\hyperlink{emailnote}{*,}}%
}

\DeclareRobustCommand{\presentdagger}{%
  \textsuperscript{\protect\hyperlink{presentnote}{\textsuperscript{\dagger,}}}%
}

\begin{document}
\title{Towards logical entanglement creation in trivalent planar architectures}
\author[1,2]{Lukas B\"{o}deker\emailstar}
\author[1,2]{Luis Colmenarez}
\author[3,4]{Sergey Blinov}
\author[5]{Ants Remm\presentdagger}
\author[5]{Simon Gustavsson\presentdagger}
\author[1,2]{Markus~M\"{u}ller}

\affil[1]{Institute for Theoretical Nanoelectronics (PGI-2), Forschungszentrum J\"{u}lich, 52428 J\"{u}lich, Germany}
\affil[2]{Institute for Quantum Information, RWTH Aachen University, 52056 Aachen, Germany}
\affil[3]{Department of Applied Physics, Yale University, New Haven, Connecticut 06511, USA}
\affil[4]{Yale Quantum Institute, Yale University, New Haven, Connecticut 06511, USA}
\affil[5]{Atlantic Quantum, Cambridge, MA 02139, USA}
\date{\today}

\twocolumn[
\begin{@twocolumnfalse}

\maketitle

\begin{abstract}
Low-overhead quantum error-correction schemes are essential for enabling quantum computation on registers containing multiple logical qubits. For planar architectures with limited nearest-neighbor qubit connectivity, the surface code has emerged as the leading paradigm. Recent theoretical and experimental work has shown that a physical-qubit connectivity of degree three is sufficient to implement fault-tolerant quantum error correction. In this work, we study lattice surgery in the context of such trivalent architectures and introduce scalable circuit constructions to implement it. Compared with the four-valent measurement scheme, the trivalent lattice-surgery protocol reduces the required resources by $\mathcal{O}(d)$ qubits out of a total qubit count of $\mathcal{O}(d^2)$ and by $\mathcal{O}(d)$ two-qubit gates out of a total two-qubit gate count of $\mathcal{O}(d^3)$. We benchmark the logical fidelity of both lattice-surgery schemes in terms of experimentally realistic simulations targeting an implementation with a fluxonium qubit based architecture and find a potential improvement of up to $\approx25\%$ for distance-three. These results open a way for scalable planar trivalent qubit architectures to host a surface-code-based logical quantum processor.
\end{abstract}

\vspace{0.5em}

\end{@twocolumnfalse}
]

\begingroup
\renewcommand{\thefootnote}{\fnsymbol{footnote}}
\footnotetext[1]{%
  \hypertarget{emailnote}{}%
  \href{mailto:l.boedeker@fz-juelich.de}{l.boedeker@fz-juelich.de}%
}
\footnotetext[2]{%
  \hypertarget{presentnote}{}%
  Present address: Google Quantum AI, Cambridge, MA 02142, United States%
}
\endgroup
\section{Introduction}
Executing useful quantum algorithms that outperform their classical counterparts requires quantum error correction (QEC) to protect quantum information and therefore yield reliable computations~\cite{dalzell2023quantum}. 
Recent experimental advances~\cite{Sundaresan23,Ryan_Anderson2024,Zhang2026,hetenyi_creating_2024,bluvstein2024logical,acharya2024quantum,wang2026,Besedin2026} have demonstrated the feasibility of running QEC schemes and subroutines on various near-term quantum devices, yet several challenges remain. 
First, the noise level of the physical components must be sufficiently low for QEC to reduce the error rate of logical qubits encoded in an error-correcting code~\cite{Dennis2002}. 
Second, arbitrary logical operations must be implemented in a fault-tolerant (FT) manner, which introduces additional overhead~\cite{EastinKnillTheo}. 
In particular, for solid-state qubit architectures, constraints on qubit connectivity (valency) play a central role in determining achievable error rates, control complexity, and scalability. 
Lower valency per physical qubit aligns better with the hardware constraints of leading 2D planar quantum platforms, and in particular superconducting qubit architectures~\cite{Sundaresan23,eickbusch2024demonstrating,acharya2024quantum,wang2026,hetenyi_creating_2024,Zhang2026,Besedin2026,lacroix_2025,lin2026surfacecodelogicaloperations,rosenfeld2025magic}.

A widely used QEC scheme is the surface code, a planar stabilizer code~\cite{Gottesman1999} with a relatively high error threshold for circuit-level noise, below which logical errors can be suppressed~\cite{Kitaev1997,Dennis2002}. 
Its conventional implementation relies on measuring stabilizers acting on four qubits (or two at boundaries), which requires ancilla qubits coupled to all qubits in the stabilizer support. 
As a result, bulk qubits must interact with four nearest neighbors on a square lattice~\cite{Fowler2012}, imposing demanding connectivity requirements on the hardware. 
The same circuitry enables lattice surgery, which realizes entangling logical operations between encoded qubits and is essential for universal fault-tolerant quantum computation~\cite{Horsman_2012,litinski2019game}. 

Recently, McEwen, Bacon, and Gidney introduced a trivalent measurement scheme that relaxes these connectivity constraints by allowing stabilizer measurements using ancilla qubits that interact with only three neighbors~\cite{mcewen2023relaxing}. 
This approach reduces hardware requirements while preserving compatibility with surface-code QEC. 
The scheme has since been extended to handle fabrication defects~\cite{higgott2025handlingfabricationdefectshexgrid}, compared across different lattice geometries~\cite{Benito2025comparativestudyof,vezvaee2025surface}, generalized to other codes such as color codes and LDPC codes~\cite{gidney2023new,shaw2024lowering,shaw2026optimising}, and experimentally demonstrated~\cite{eickbusch2024demonstrating}. 
%However, application of the trivalent circuit design to logical operations, in particular lattice surgery, has not been explored in the context of surface codes.
We study the application of the trivalent circuit design for lattice surgery operations and benchmark them for experimentally relevant settings. Similar trivalent lattice surgery schemes have been used in Refs.~\cite{low2026denser,hirai2026no} as components of larger fault-tolerant schemes. By contrast, we focus on a direct comparison with the more conventional four-valent circuit design and evaluate both approaches under realistic conditions, including those relevant to fluxonium-qubit architectures.
In this manuscript, we develop and analyze a trivalent lattice-surgery scheme for the rotated surface code hosted on planar qubit layouts. 
We first present the trivalent circuit construction and map it to experimentally relevant implementations, emphasizing its suitability for near-term hardware. 
We then benchmark the surface code under both trivalent and four-valent measurement schemes at small code distances using an experimentally motivated circuit-level noise model, with a focus on superconducting qubit architectures. 
In this setting, we reproduce the quantum memory results of Ref.~\cite{mcewen2023relaxing} and identify trade-offs arising from reduced connectivity and the experimentally motivated noise model. 

Building on this, we then introduce a lattice-surgery protocol that operates without requiring additional data qubits in the intermediate region between logical patches. 
This reduces the qubit overhead compared to four-valent implementations~\cite{Horsman_2012} while maintaining modularity of logical operations. 
Finally, we show that the trivalent layout provides a direct advantage over four-valent lattice surgery by reducing the number of required gates by a subleading factor for growing system sizes. Together with expected hardware improvements due to the trivalent layout, this leads to an anticipated improved logical performance of the lattice surgery.

The manuscript is organized as follows. 
In~\cref{sec:hex_op}, we present the trivalent circuit design and its operational principles. 
In~\cref{sec:surgery}, we introduce the trivalent lattice-surgery protocol and compare its performance to the four-valent approach. 
Finally, in~\cref{sec:conclusion}, we discuss the implications of our results for superconducting architectures, with particular emphasis on fluxonium qubits.

\section{Operation of the surface code on a hexagonal grid}
\label{sec:hex_op}

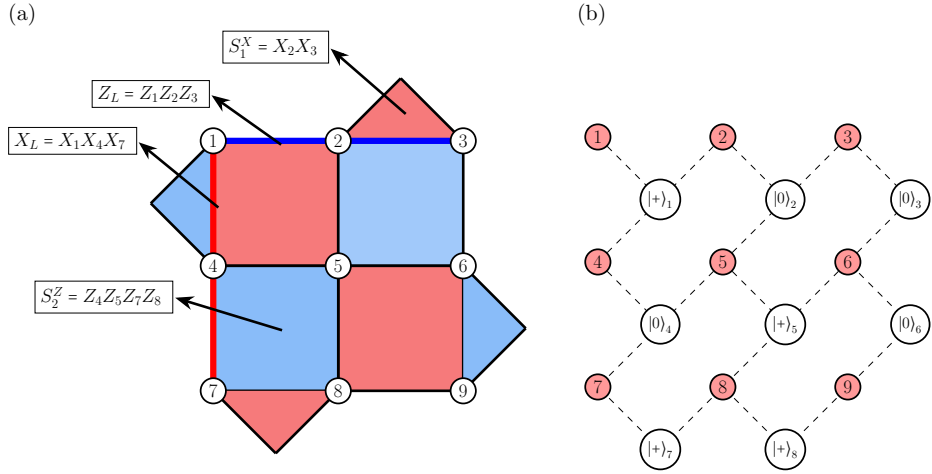
\begin{figure*}[!t]
    \centering
    \resizebox{1.5\columnwidth}{!}{
    \begin{tabular}{@{}c@{\hspace{0.8cm}}c@{}}

%------------------------------------------------
% Left figure
%------------------------------------------------
\multicolumn{1}{@{}l@{}}{\scalebox{1}{(a)}}&\multicolumn{1}{@{}l@{}}{\scalebox{1}{(b)}}\\[0.05cm]
\begin{tikzpicture}[scale=0.5,transform shape]
  \tikzset{
    qubit/.style={circle,thick,draw,fill=white}
  }

  % coords
  \coordinate (c3) at (0,0);
  \coordinate (c1) at (0,4);
  \coordinate (c4) at (4,0);
  \coordinate (c2) at (4,4);
  \coordinate (c0) at (2,2);
  \coordinate (c6) at (6,2);
  \coordinate (c5) at (2,-2);

  \coordinate (c7) at (8,4);
  \coordinate (c8) at (8,0);
  \coordinate (c9) at (-2,2);
  \coordinate (c10) at (6,-2);

  \coordinate (c11) at (0,-4);
  \coordinate (c12) at (4,-4);
  \coordinate (c13) at (8,-4);
  \coordinate (c14) at (10,-2);
  \coordinate (c15) at (2,-6);
  \coordinate (c16) at (6,6);

  % stabilizers
  \path[fill=stimBlue]
    (c1) -- (c3) -- (c9) -- cycle;

  \draw[very thick,fill=stimRed]
    (c1) -- (c2) -- (c4) -- (c3) -- cycle;

  \draw[very thick,fill=stimBlue!80]
    (c2) -- (c7) -- (c8) -- (c4) -- cycle;

  \draw[very thick,fill=stimRed]
    (c8) -- (c4) -- (c12) -- (c13) -- cycle;

  \draw[very thick,fill=stimBlue]
    (c3) -- (c4) -- (c12) -- (c11) -- cycle;

  \path[fill=stimBlue]
    (c8) -- (c14) -- (c13) -- cycle;

  \path[fill=stimRed]
    (c2) -- (c16) -- (c7) -- cycle;

  \path[fill=stimRed]
    (c11) -- (c15) -- (c12) -- cycle;

  % logical operators
  \draw[line width=1mm,color=blue]
    (c1) -- (c2) -- (c7);

  \draw[line width=1mm,color=red]
    (c1) -- (c3) -- (c11);

  % boundaries
  \draw[very thick] (c3) -- (c9);
  \draw[very thick] (c1) -- (c9);

  \draw[very thick] (c14) -- (c13);
  \draw[very thick] (c14) -- (c8);

  \draw[very thick] (c15) -- (c11);
  \draw[very thick] (c15) -- (c12);

  \draw[very thick] (c16) -- (c7);
  \draw[very thick] (c16) -- (c2);

  % data qubits
  \node[qubit] at (c1) {\LARGE 1};
  \node[qubit] at (c2) {\LARGE 2};
  \node[qubit] at (c3) {\LARGE 4};
  \node[qubit] at (c4) {\LARGE 5};
  \node[qubit] at (c7) {\LARGE 3};
  \node[qubit] at (c8) {\LARGE 6};
  \node[qubit] at (c11) {\LARGE 7};
  \node[qubit] at (c12) {\LARGE 8};
  \node[qubit] at (c13) {\LARGE 9};

  % annotations
  \node[draw,align=left,inner sep=5pt] (T1) at (2,7) {%
    \LARGE $S^X_1=X_2X_3$
  };
  \coordinate (p1) at (6,5);
  \draw[-Stealth,very thick,shorten <=-3pt,shorten >=1pt]
    (p1) -- (T1.east);

  \node[draw,align=left,inner sep=5pt] (T2) at (-2,5.5) {%
    \LARGE $Z_L=Z_1Z_2Z_3$
  };
  \coordinate (p2) at (2,4);
  \draw[-Stealth,very thick,shorten <=-3pt,shorten >=1pt]
    (p2) -- (T2.east);

  \node[draw,align=left,inner sep=5pt] (T3) at (-4.5,4) {%
    \LARGE $X_L=X_1X_4X_7$
  };
  \coordinate (p3) at (0,2);
  \draw[-Stealth,very thick,shorten <=-3pt,shorten >=1pt]
    (p3) -- (T3.east);

  \node[draw,align=left,inner sep=5pt] (T4) at (-3.5,-1) {%
    \LARGE $S^Z_2=Z_4Z_5Z_7Z_8$
  };
  \draw[-Stealth,very thick,shorten <=-3pt,shorten >=1pt]
    (c5) -- (T4.east);

\end{tikzpicture}

&

%------------------------------------------------
% Right figure
%------------------------------------------------
\raisebox{-0.25cm}{
\begin{tikzpicture}[scale=0.5,transform shape]
  \tikzset{
    qubit/.style={circle,thick,draw,fill=white}
  }

  % coords
  \coordinate (c3) at (0,0);
  \coordinate (c1) at (0,4);
  \coordinate (c4) at (4,0);
  \coordinate (c2) at (4,4);
  \coordinate (c0) at (2,2);
  \coordinate (c6) at (6,2);
  \coordinate (c5) at (2,-2);

  \coordinate (c7) at (8,4);
  \coordinate (c8) at (8,0);
  \coordinate (c9) at (10,2);
  \coordinate (c10) at (6,-2);

  \coordinate (c11) at (0,-4);
  \coordinate (c12) at (4,-4);
  \coordinate (c13) at (8,-4);
  \coordinate (c14) at (10,-2);
  \coordinate (c15) at (2,-6);
  \coordinate (c16) at (6,-6);

  % connectivity
  \draw[dashed] (c1) -- (c0);
  \draw[dashed] (c3) -- (c0);
  \draw[dashed] (c3) -- (c5);
  \draw[dashed] (c2) -- (c0);
  \draw[dashed] (c2) -- (c6);
  \draw[dashed] (c4) -- (c6);
  \draw[dashed] (c7) -- (c6);
  \draw[dashed] (c4) -- (c5);
  \draw[dashed] (c7) -- (c9);
  \draw[dashed] (c8) -- (c9);
  \draw[dashed] (c4) -- (c10);
  \draw[dashed] (c8) -- (c10);
  \draw[dashed] (c5) -- (c11);
  \draw[dashed] (c10) -- (c12);
  \draw[dashed] (c14) -- (c13);
  \draw[dashed] (c14) -- (c8);
  \draw[dashed] (c15) -- (c11);
  \draw[dashed] (c15) -- (c12);
  \draw[dashed] (c16) -- (c13);
  \draw[dashed] (c16) -- (c12);

  % data qubits
  \node[qubit,fill=red!40] at (c1) {\LARGE 1};
  \node[qubit,fill=red!40] at (c2) {\LARGE 2};
  \node[qubit,fill=red!40] at (c3) {\LARGE 4};
  \node[qubit,fill=red!40] at (c4) {\LARGE 5};
  \node[qubit,fill=red!40] at (c7) {\LARGE 3};
  \node[qubit,fill=red!40] at (c8) {\LARGE 6};
  \node[qubit,fill=red!40] at (c11) {\LARGE 7};
  \node[qubit,fill=red!40] at (c12) {\LARGE 8};
  \node[qubit,fill=red!40] at (c13) {\LARGE 9};

  % ancilla qubits
  \node[qubit] at (c0) {\Large$\ket{+}_1$};
  \node[qubit] at (c6) {\Large$\ket{0}_2$};
  \node[qubit] at (c5) {\Large$\ket{0}_4$};
  \node[qubit] at (c9) {\Large $\ket{0}_3$};
  \node[qubit] at (c10) {\Large$\ket{+}_5$};
  \node[qubit] at (c14) {\Large$\ket{0}_6$};
  \node[qubit] at (c15) {\Large$\ket{+}_7$};
  \node[qubit] at (c16) {\Large$\ket{+}_8$};

\end{tikzpicture}
}
\end{tabular}}
    \caption{Illustrations of a distance $d=3$ rotated surface code or surface-17 patch. In (a) the definition of the code in terms of its stabilizer generators and logical operators is indicated. Generally, red stands for $X$-type operators and blue stands for $Z$-type operators. The support of the stabilizer generators corresponds to corner qubits of the colored plaquette, while the logical operators have support on the colored lines. In (b) the qubit layout for the trivalent surface code is given with the same data-qubit numbering. A valid initialization state of the ancilla qubits is indicated.}
    \label{fig:s17}
\end{figure*}

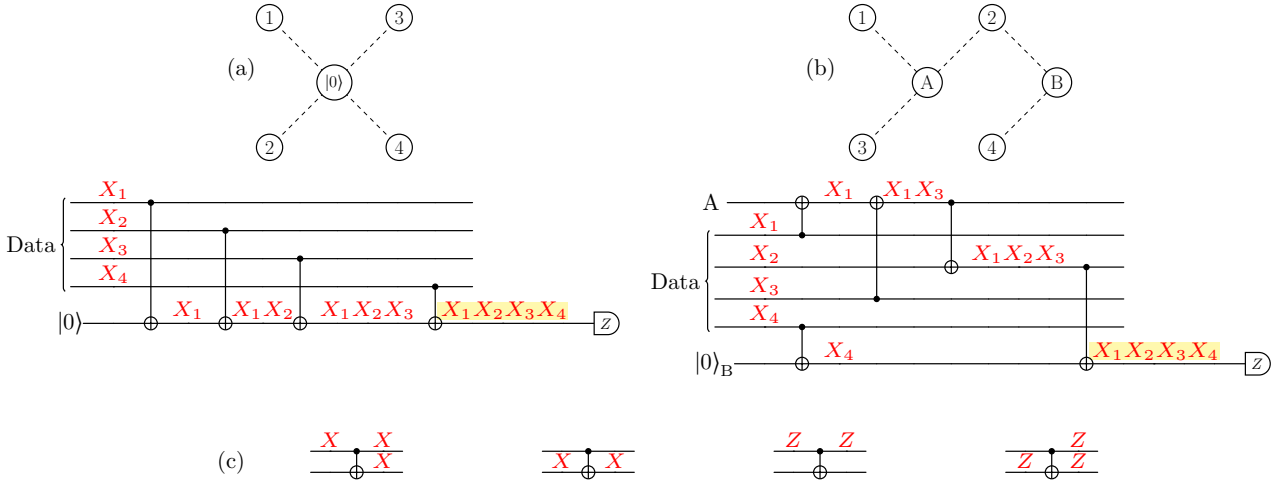
\begin{figure*}[!t]
    \centering
    \resizebox{2\columnwidth}{!}{%

\newcommand{\largetextP}[1]{\scalebox{1.5}{$#1$}}
\begin{tabular}{c}

\begin{tabular}{@{}c@{\hspace{0.5cm}}c@{}}

%---------------- Panel (a) ----------------%
\begin{tabular}[t]{@{}c@{}}

\llap{\raisebox{2cm}{\scalebox{1.5}{(a)}}}%
\begin{tikzpicture}[scale=0.75,transform shape]
  \tikzset{
    qubit/.style={circle,thick,draw,fill=white}
  }

  \coordinate (c3) at (0,0);
  \coordinate (c1) at (0,4);
  \coordinate (c4) at (4,0);
  \coordinate (c2) at (4,4);
  \coordinate (c0) at (2,2);

  \draw[dashed] (c1) -- (c0);
  \draw[dashed] (c3) -- (c0);
  \draw[dashed] (c2) -- (c0);
  \draw[dashed] (c4) -- (c0);

  \node[qubit] at (c1) {\LARGE 1};
  \node[qubit] at (c2) {\LARGE 3};
  \node[qubit] at (c3) {\LARGE 2};
  \node[qubit] at (c4) {\LARGE 4};
  \node[qubit] at (c0) {\Large$\ket{0}$};
\end{tikzpicture}
\\[0.75em]

\hspace*{-1.0cm}%
\begin{quantikz}[column sep=0.7cm]
  \lstick[4]{\largetextP{\text{Data}}}
      & \opname{X_1}
      & \ctrl{4}
      &&&&&&& &&\\
  & \opname{X_2}
      &&& \ctrl{3}
      &&&&& &&\\
  & \opname{X_3}
      &&&&& \ctrl{2}
      &&& &&\\
  & \opname{X_4}
      &&&&&&& &&\ctrl{1}
      & \\
  \largetextP{\ket{0}}
      && \targ{}
      & \opname{X_1}
      & \targ{}
      & \opname{X_1X_2}
      & \targ{}
      & &\opname{X_1X_2X_3}&
      & \targ{}
      && \highlightop{\,X_1X_2X_3X_4}&
      && \meterD{Z}
\end{quantikz}

\end{tabular}

&

%---------------- Panel (b) ----------------%
\begin{tabular}[t]{@{}c@{}}

\llap{\raisebox{2cm}{\scalebox{1.5}{(b)}}\hspace{0.3cm}}%
\begin{tikzpicture}[scale=0.75,transform shape]
  \tikzset{
    qubit/.style={circle,thick,draw,fill=white}
  }

  \coordinate (c3) at (0,0);
  \coordinate (c1) at (0,4);
  \coordinate (c4) at (4,0);
  \coordinate (c2) at (4,4);
  \coordinate (c0) at (2,2);
  \coordinate (c5) at (6,2);

  \draw[dashed] (c1) -- (c0);
  \draw[dashed] (c3) -- (c0);
  \draw[dashed] (c2) -- (c0);
  \draw[dashed] (c2) -- (c5);
  \draw[dashed] (c4) -- (c5);

  \node[qubit] at (c1) {\LARGE 1};
  \node[qubit] at (c2) {\LARGE 2};
  \node[qubit] at (c3) {\LARGE 3};
  \node[qubit] at (c4) {\LARGE 4};
  \node[qubit] at (c5) {\LARGE B};
  \node[qubit] at (c0) {\LARGE A};
\end{tikzpicture}
\\[0.75em]

\begin{quantikz}[column sep=0.7cm]
  \largetextP{\text{A }}
      && \targ{}
      & \opname{X_1}
      & \targ{}
      & \opname{X_1X_3}
      & \ctrl{2}
      &&& &&\\
  \lstick[4]{\largetextP{\text{Data}}}
      & \opname{X_1}
      & \ctrl{-1}
      &&&&&&&&& \\
  & \opname{X_2}
      &&&&& \targ{}
      & &\opname{X_1X_2X_3}
      && \ctrl{3}
      & \\
  & \opname{X_3}
      &&& \ctrl{-3}
      &&&&&& &\\
  & \opname{X_4}
      & \ctrl{1}
      &&&&&&&& &\\
  \largetextP{\ket{0}_{\text{B}}}
      && \targ{}
      & \opname{X_4}
      &&&&&& &\targ{}
      && \highlightop{\,X_1X_2X_3X_4}&
      && \meterD{Z}
\end{quantikz}

\end{tabular}

\end{tabular}

\\[15em]

% ------------------------------------------------------------
% Lower block: new panel (a), four small circuits
% ------------------------------------------------------------
\begin{tabular}{c@{\hspace{1.4cm}}c@{\hspace{3cm}}c@{\hspace{3cm}}c@{\hspace{3cm}}c}
\scalebox{1.5}{(c)}
&
\begin{quantikz}[column sep=0.45cm, row sep=0.25cm]
& \opname{X} & \ctrl{1} &\opname{X}& \\
  &              & \targ{}  & \opname{X}&
\end{quantikz}
&
\begin{quantikz}[column sep=0.45cm, row sep=0.25cm]
& & \ctrl{1} && \\
  &    \opname{X}           & \targ{}  & \opname{X}&
\end{quantikz}
&
\begin{quantikz}[column sep=0.45cm, row sep=0.25cm]
& \opname{Z} & \ctrl{1} &\opname{Z}& \\
  &              & \targ{}  & &
\end{quantikz}
&
\begin{quantikz}[column sep=0.45cm, row sep=0.25cm]
&  & \ctrl{1} &\opname{Z}& \\
  &     \opname{Z}         & \targ{}  & \opname{Z}&
\end{quantikz}
\end{tabular}

\end{tabular}

}
    \caption{
     Comparison of the measurement circuits of single weight-four $Z$ plaquette stabilizer parity between the four-valent measurement scheme (a) and the trivalent measurement scheme (b). The respective qubit layouts, including their valencies (upper row) and the circuitry are shown (lower row). In the two measurement circuits, the propagation of potential single $X$ errors incident on the data qubits is tracked to show explicitly how the respective error is recorded by measuring the ancilla. The notation of $X_i$ being present shall be understood as a potential error on each of the data qubits. The ancilla initialization and measurement for the trivalent measurement scheme is adapted compared to the bulk~\cref{fig:trivalent_readout_cycle_s17} to demonstrate the stabilizer measurement in isolation. The highlighted operator before the $Z$ measurement of the ancilla qubit indicates which error parity is measured.  In (c) we summarize the error propagation rules through a CNOT gate.
     }
\label{fig:trivalent_readout_cycle_plaquette}
\end{figure*}

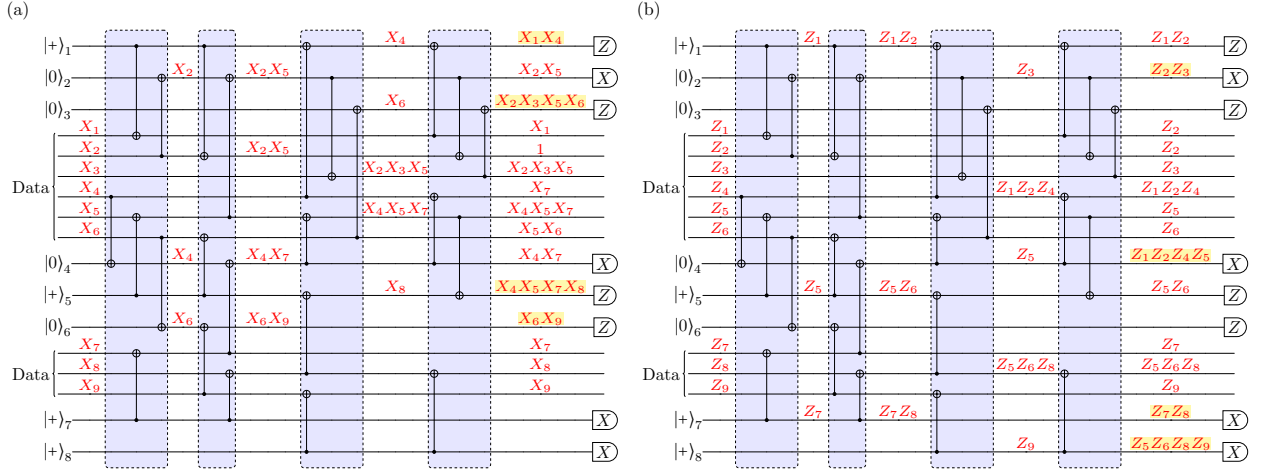
\begin{figure*}[!t]
    \centering
\resizebox{2\columnwidth}{!}{\newcommand{\opnameL}[1]{%
  \wire["\scalebox{2.5}{\textcolor{red}{$#1$}}"{above,pos=1}]{a}%
}

\newcommand{\highlightopL}[1]{%
  \wire[
    "\colorbox{yellow!40}{\scalebox{2.5}{\textcolor{red}{$#1$}}}"
    {above,pos=1}
  ]{a}%
}

\begin{tabular}{@{}c@{\hspace{0.4cm}}c@{}}

%==================== Panel (a) ====================%
\begin{tabular}[t]{@{}c@{}}
\multicolumn{1}{@{}l@{}}{\scalebox{1.5}{(a)}}
\\[0.4em]
\scalebox{0.72}{
\begin{quantikz}[column sep=0.7cm]
\largetext{\ket{+}_1}&&\boxmark{3}&\ctrl{3}&&&\ctrl{4}\boxmark{2}&&&&&\targ{}\boxmark{3}&&&&\opnameL{X_4}&&\targ{}\boxmark{3}&&&&&\highlightopL{X_1X_4}&&&\meterD{\largetext{Z}}
  \\
\largetext{\ket{0}_2}&&&&\targ{}&\opnameL{X_2}&&\targ{}&&\opnameL{X_2X_5}&&&\ctrl{4}&&&&&&\ctrl{3}&&&&\opnameL{X_2X_5}&&&\meterD{\largetext{X}}
  \\
\largetext{\ket{0}_3}&&&&&&&&&&&&&\targ{}&&\opnameL{X_6}&&&&\targ{}&&&\highlightopL{X_2X_3X_5X_6}&&&\meterD{\largetext{Z}}
  \\
\lstick[6]{\largetext{\text{Data}}}&\opnameL{X_1}&&\targ{}&&&&&&&&&&&&&&\ctrl{-3}&&&&&\opnameL{X_1}&&&
  \\
&\opnameL{X_2}&&&\ctrl{-3}&&\targ{}&&&\opnameL{X_2X_5}&&&&&&&&&\targ{}&&&&\opnameL{1}&&&
  \\
&\opnameL{X_3}&&&&&&&&&&&\targ{}&&&\opnameL{X_2X_3X_5}&&&&\ctrl{-3}&&&\opnameL{X_2X_3X_5}&&&
  \\
&\opnameL{X_4}&\ctrl{3}&&&&&&&&&\ctrl{-6}&&&&&&\targ{}&&&&&\opnameL{X_7}&&&
  \\
&\opnameL{X_5}&&\targ{}&&&&\ctrl{-6}&&&&\targ{}&&&&\opnameL{X_4X_5X_7}&&&\ctrl{3}&&&&\opnameL{X_4X_5X_7}&&&
  \\
&\opnameL{X_6}&&&\ctrl{3}&&\targ{}&&&&&&&\ctrl{-6}&&&&&&&&&\opnameL{X_5X_6}&&&
  \\
\largetext{\ket{0}_4}&&\targ{}&&&\opnameL{X_4}&&\targ{}&&\opnameL{X_4X_7}&&\ctrl{-2}&&&&&&\ctrl{-3}&&&&&\opnameL{X_4X_7}&&&\meterD{\largetext{X}}
  \\
\largetext{\ket{+}_5}&&&\ctrl{-3}&&&\ctrl{-2}&&&&&\targ{}&&&&\opnameL{X_8}&&&\targ{}&&&&\highlightopL{X_4X_5X_7X_8}&&&\meterD{\largetext{Z}}
  \\
\largetext{\ket{0}_6}&&&&\targ{}&\opnameL{X_6}&\targ{}&&&\opnameL{X_6X_9}&&&&&&&&&&&&&\highlightopL{X_6X_9}&&&\meterD{\largetext{Z}}
  \\
\lstick[3]{\largetext{\text{Data}}}&\opnameL{X_7}&&\targ{}&&&&\ctrl{-3}&&&&&&&&&&&&&&&\opnameL{X_7}&&&
  \\
&\opnameL{X_8}&&&&&&\targ{}&&&&\ctrl{-3}&&&&&&\targ{}&&&&&\opnameL{X_8}&&&
  \\
&\opnameL{X_9}&&&&&\ctrl{-3}&&&&&\targ{}&&&&&&&&&&&\opnameL{X_9}&&&
  \\
\largetext{\ket{+}_7}&&&\ctrl{-3}&&&&\ctrl{-2}&&&&&&&&&&&&&&&&&&\meterD{\largetext{X}}
  \\
\largetext{\ket{+}_8}&&&&&&&&&&&\ctrl{-2}&&&&&&\ctrl{-3}&&&&&&&&\meterD{\largetext{X}}
\end{quantikz}
}
\end{tabular}

&

%==================== Panel (b) ====================%
\begin{tabular}[t]{@{}c@{}}
\multicolumn{1}{@{}l@{}}{\scalebox{1.5}{(b)}}
\\[0.4em]
\scalebox{0.72}{
\begin{quantikz}[column sep=0.7cm]
\largetext{\ket{+}_1}&&\boxmark{3}&\ctrl{3}&&\opnameL{Z_1}&\ctrl{4}\boxmark{2}&&&\opnameL{Z_1Z_2}&&\targ{}\boxmark{3}&&&&&&\targ{}\boxmark{3}&&&&&\opnameL{Z_1Z_2}&&&\meterD{\largetext{Z}}
  \\
\largetext{\ket{0}_2}&&&&\targ{}&&&\targ{}&&&&&\ctrl{4}&&&\opnameL{Z_3}&&&\ctrl{3}&&&&\highlightopL{Z_2Z_3}&&&\meterD{\largetext{X}}
  \\
\largetext{\ket{0}_3}&&&&&&&&&&&&&\targ{}&&&&&&\targ{}&&&&&&\meterD{\largetext{Z}}
  \\
\lstick[6]{\largetext{\text{Data}}}&\opnameL{Z_1}&&\targ{}&&&&&&&&&&&&&&\ctrl{-3}&&&&&\opnameL{Z_2}&&&
  \\
&\opnameL{Z_2}&&&\ctrl{-3}&&\targ{}&&&&&&&&&&&&\targ{}&&&&\opnameL{Z_2}&&&
  \\
&\opnameL{Z_3}&&&&&&&&&&&\targ{}&&&&&&&\ctrl{-3}&&&\opnameL{Z_3}&&&
  \\
&\opnameL{Z_4}&\ctrl{3}&&&&&&&&&\ctrl{-6}&&&&\opnameL{Z_1Z_2Z_4}&&\targ{}&&&&&\opnameL{Z_1Z_2Z_4}&&&
  \\
&\opnameL{Z_5}&&\targ{}&&&&\ctrl{-6}&&&&\targ{}&&&&&&&\ctrl{3}&&&&\opnameL{Z_5}&&&
  \\
&\opnameL{Z_6}&&&\ctrl{3}&&\targ{}&&&&&&&\ctrl{-6}&&&&&&&&&\opnameL{Z_6}&&&
  \\
\largetext{\ket{0}_4}&&\targ{}&&&&&\targ{}&&&&\ctrl{-2}&&&&\opnameL{Z_5}&&\ctrl{-3}&&&&&\highlightopL{Z_1Z_2Z_4Z_5}&&&\meterD{\largetext{X}}
  \\
\largetext{\ket{+}_5}&&&\ctrl{-3}&&\opnameL{Z_5}&\ctrl{-2}&&&\opnameL{Z_5Z_6}&&\targ{}&&&&&&&\targ{}&&&&\opnameL{Z_5Z_6}&&&\meterD{\largetext{Z}}
  \\
\largetext{\ket{0}_6}&&&&\targ{}&&\targ{}&&&&&&&&&&&&&&&&&&&\meterD{\largetext{Z}}
  \\
\lstick[3]{\largetext{\text{Data}}}&\opnameL{Z_7}&&\targ{}&&&&\ctrl{-3}&&&&&&&&&&&&&&&\opnameL{Z_7}&&&
  \\
&\opnameL{Z_8}&&&&&&\targ{}&&&&\ctrl{-3}&&&&\opnameL{Z_5Z_6Z_8}&&\targ{}&&&&&\opnameL{Z_5Z_6Z_8}&&&
  \\
&\opnameL{Z_9}&&&&&\ctrl{-3}&&&&&\targ{}&&&&&&&&&&&\opnameL{Z_9}&&&
  \\
\largetext{\ket{+}_7}&&&\ctrl{-3}&&\opnameL{Z_7}&&\ctrl{-2}&&\opnameL{Z_7Z_8}&&&&&&&&&&&&&\highlightopL{Z_7Z_8}&&&\meterD{\largetext{X}}
  \\
\largetext{\ket{+}_8}&&&&&&&&&&&\ctrl{-2}&&&&\opnameL{Z_9}&&\ctrl{-3}&&&&&\highlightopL{Z_5Z_6Z_8Z_9}&&&\meterD{\largetext{X}}
\end{quantikz}
}
\end{tabular}

\end{tabular}}
    \caption{Circuitry for conducting a single QEC cycle for the trivalent measurement scheme. The next readout cycle would correspond to the execution of an altered circuit and is given in~\cref{fig:memory_circuit}. The panels (a) and (b) show the same circuit, but demonstrate the measurement of $Z$-type stabilizers and $X$-type stabilizers by showing the propagation of $X$ and $Z$-type data-qubit errors.
    After initializing the ancilla qubits, entangling gates are conducted in four time steps. Finally, the ancilla qubits are measured in the appropriate basis to obtain the stabilizer value. The $X$ and $Z$ stabilizers are measured in an intertwined manner, but still maintaining the same gate count as in the four-valent circuitry. The qubit labeling is consistent with the layout in~\cref{fig:s17}~(b)}
    \label{fig:trivalent_readout_cycle_s17}
\end{figure*}

We focus on a two-dimensional planar logical quantum processor based on the rotated surface code~\cite{Bombin2007}, with the goal of independently controlling logical qubits while enabling entanglement generation via lattice surgery~\cite{Horsman_2012}. 
The rotated surface code consists of weight-four plaquette operators alternating between $X$- and $Z$-type on the lattice. An example for code distance $d=3$ is shown in~\cref{fig:s17}. %At the boundaries, every second plaquette is reduced to a weight-two operator (triangles in~\cref{fig:s17}), while the remaining plaquettes are removed. 
Logical state initialization of eigenstates of $X_L$ or $Z_L$, %i.e., $\{\ket{0}_L,\ket{1}_L,\ket{+}_L,\ket{-}_L\}$
can be performed fault-tolerantly by performing $\mathcal{O}(d)$ rounds of stabilizer measurements~\cite{Dennis2002}. Additional rounds of $X$ and $Z$ stabilizer measurements (QEC cycles) can be executed to suppress the probability of logical errors during idling, provided the underlying physical error rates are sufficiently low. 
Other logical state preparations, such as $Y$-basis eigenstates, require adapted schemes~\cite{gidney2024inplace}, while the preparation of magic states, required for universal logical gates, incurs substantially higher overhead~\cite{gidney2024magicstatecultivationgrowing,litinski2019magic,rosenfeld2025magic,Rodriguez2025}. Once available, these magic states enable non-Clifford operations via logical state injection~\cite{jozsa2006,litinski2019game,Kim2026}. 

Beyond state preparation, the creation of logical entanglement and two-qubit logical gates can be achieved via lattice surgery between two logical surface-code patches. In this procedure, the values of joint logical operators are determined by measuring low-weight parity operators between the patches~\cite{Horsman_2012}. These parity operators can be viewed as stabilizers of weight four with partial support on one or both logical qubits, and potentially on additional data qubits hosted between them. This is visualized for different qubit and connectivity layouts in~\cref{fig:modularity}. The initial projective measurement of these surgery stabilizers effectively merges the two code patches. Based on this operation, single logical qubit gates can be injected, or with access to a third surface code, a logical entangling gate can be performed with two merging operations~\cite{Horsman_2012}.
In summary, the measurement of stabilizer generators lies at the heart of operating a logical quantum processor based on planar surface codes. 

Accordingly, it is critical to optimize this subroutine which is executed repeatedly throughout any logical quantum computation.

\subsection{Measuring a plaquette operator}

We now explain how the required connectivity of the physical qubits can be reduced for measuring surface-code stabilizers, following Ref.~\cite{mcewen2023relaxing}. 
To illustrate this, we first focus on the circuit for measuring a single weight-four stabilizer operator, without yet considering the larger surface-code embedding. 
In the four-valent approach, such a stabilizer is measured using a measurement ancilla qubit coupled to four data qubits, as shown in~\cref{fig:trivalent_readout_cycle_plaquette}(a). 
The measurement is performed by applying four successive entangling gates between the ancilla and each of the four data qubits on which the stabilizer has support. 
When embedded in the surface code, the ordering of these entangling gates must be chosen carefully so that any error propagation from the ancilla to the data qubits remains correctable~\cite{Dennis2002,Tomita2014}. 
Alternatively, Ref.~\cite{mcewen2023relaxing} proposes a circuit design to measure the same stabilizer that requires each qubit to interact with at most three neighbors. 
We refer to this paradigm as the \emph{trivalent circuit design} or \emph{trivalent readout schedule}, an example layout of which for a single plaquette is shown in~\cref{fig:trivalent_readout_cycle_plaquette}(b).

We wish to provide a complementary explanation of \emph{how} a surface-code stabilizer of weight four is measured in the trivalent measurement scheme by considering error propagation from data qubits to ancillary qubits, where errors are ultimately detected in terms of a violated stabilizer. 
To do so, we compare the four-valent circuit design with the trivalent one in~\cref{fig:trivalent_readout_cycle_plaquette}. 
In the four-valent design,~\cref{fig:trivalent_readout_cycle_plaquette}(a), we consider each possible error on the data qubits that could be present before the circuit.
For a $Z$-type stabilizer, this corresponds to an error $X_i$ on data qubit $i$. This is the dual perspective of showing that the desired $Z$ type stabilizer is measured.
Any such $X$ error propagates to the ancilla qubit through the corresponding CNOT gate. As a result, the measurement outcome of the ancilla encodes the parity of $X$ errors on the data qubits. Note that this picture is the converse of the more common description that tracks the instantaneous stabilizer operators while the circuit is executed. 
The latter corresponds to a formulation in terms of \emph{detection regions}~\cite{gidney2021stim}, which we will later employ to present longer circuits. 
From the perspective of error propagation, it is essential that the same mapping of data-qubit errors $X_i$ onto the measurement ancilla also holds in the trivalent circuit design. The goal is therefore to measure the parity of these errors via the readout of the rightmost ancilla qubit~\cref{fig:trivalent_readout_cycle_plaquette}(b). However, it is not possible to couple all data qubits directly to this ancilla. Instead, additional ancilla qubits are introduced as \emph{bridge qubits} that mediate the transfer of error information.
In~\cref{fig:trivalent_readout_cycle_plaquette}(b), the central ancilla qubit A serves as such a bridge qubit. In a surface-code embedding, this qubit would be used to measure a $X$-type stabilizer. By coupling the bridge qubit A to the two data qubits 1 and 3 to its left, any $X$ errors on these qubits are first transferred onto the bridge qubit. Through subsequent gates, this error information is then propagated to the data qubit at the upper-right of the plaquette.
In this way, the parity of $X$ errors across all four data qubits is effectively mapped onto the $Z$ parity of the two data qubits on the right-hand side of the plaquette. This parity is finally extracted with the help of the rightmost ancilla qubit, whose measurement outcome yields the stabilizer value of the weight-four plaquette considered here.

\begin{figure*}[!t]
\centering
\includegraphics[width=\textwidth]{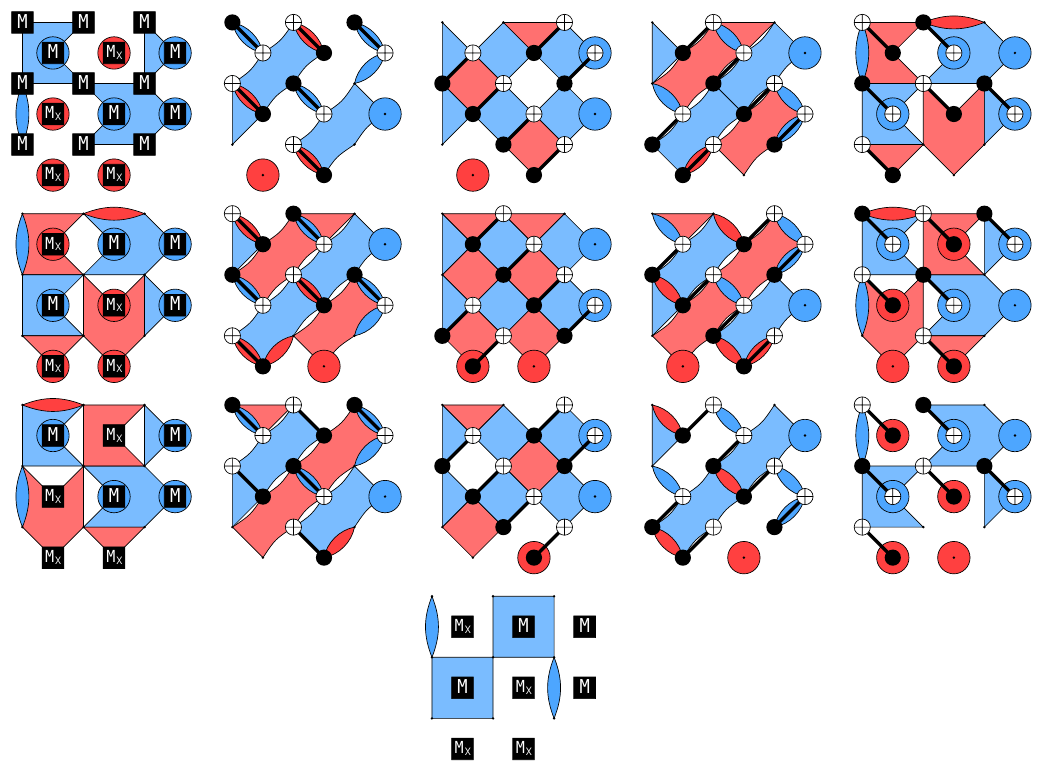}

\caption{Circuitry for initializing the $\ket{0}_L$ state of the $d=3$ rotated surface code fault tolerantly by conducting three rounds of stabilizer measurements using the trivalent circuit design. To be read from left to right and then top to bottom.}
\label{fig:memory_circuit}
\end{figure*}

A key difference of the trivalent design, compared to the four-valent one, is that errors on data qubits spread to other data qubits during a single stabilizer measurement. 
This issue is resolved by modifying the QEC cycle: in the trivalent measurement scheme, the stabilizer measurement circuit is not simply repeated, but alternated with a second type of circuit~\cite{mcewen2023relaxing}. 
This readout circuit is defined as the adjoint unitary corresponding to the original circuit, i.e., reversing the order of the CNOT gates followed by measuring the bulk ancilla qubits in the respective complementary basis. The full readout circuit for a FT initialization of $\ket{0}_L$ is shown in~\cref{fig:memory_circuit} for distance $d=3$. Here, after projectively initializing the $X$-type stabilizers, two rounds of syndrome measurement with alternating circuits are performed.

Dwelling further on the plaquette example of~\cref{fig:trivalent_readout_cycle_plaquette}, one can analyze the respective gate sequence in the context of FT circuit design. 
In the four-valent circuit~\cref{fig:trivalent_readout_cycle_plaquette}(a), an ancilla error can propagate to produce a weight-two error.
%$Z_1 Z_2$, or equivalently $Z_3 Z_4$.
To preserve fault tolerance, this plaquette must be embedded in a larger surface-code patch where the logical $Z_L$ operator can be represented along the horizontal boundary. In this configuration, the resulting error is orthogonal to the minimal support of $Z_L$ (see~\cref{fig:s17}).
In the trivalent circuit~\cref{fig:trivalent_readout_cycle_plaquette}(b), $Z$-type errors on the ancilla qubits can similarly induce weight-two data-qubit errors % $Z_1 Z_3$ or $Z_2 Z_4$
and the same requirement therefore applies: the plaquette must be embedded such that the logical $Z_L$ operator lies along the horizontal boundary. This ensures that such errors remain correctable, as they do not overlap with the minimal support of $Z_L$.

A notable difference in the single-plaquette readout is that the trivalent measurement scheme requires one additional ancilla qubit and one additional entangling gate. 
As shown in~\cref{fig:trivalent_readout_cycle_s17}, these extra resources are used in parallel to measure other stabilizers, so that, when considering the full surface code, the overall gate count and qubit overhead of the trivalent design remain the same as in the four-valent measurement scheme.

\subsection{Studying the performance of repeated QEC cycles}
\label{subsec:surf17}
\begin{figure}[!t]
    \centering
    \includegraphics[width=1\columnwidth]{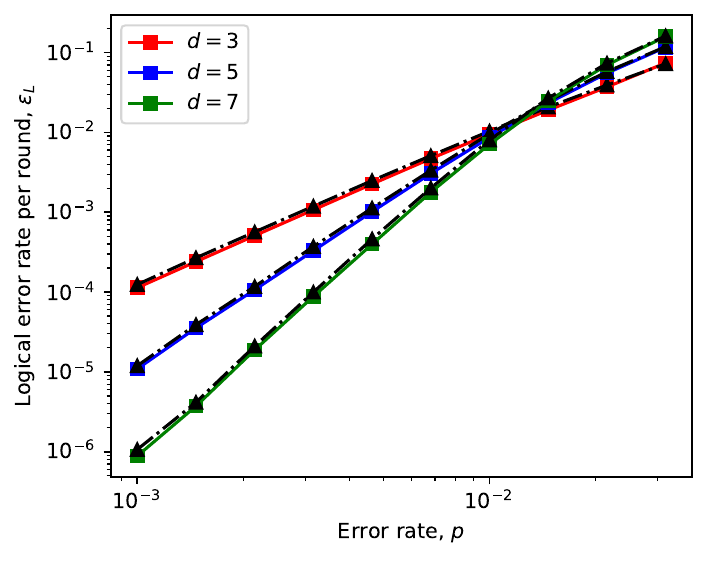}
    \caption{Comparison of the four-valent and the trivalent measurement scheme for a memory experiment of the logical $\ket{0}_L$ state. The errors occur according to a standard circuit-level noise model, outlined in App.~\ref{sec:app_noise}, where errors on two-qubit gates,  measurements, and initializations take place with a probability $p$. The logical error rate per round of stabilizer measurement is
    shown. The four-valent circuitry is indicated by the colored squares (\cbox{red}) and solid lines while the trivalent variant is given by the corresponding black triangles ($\blacktriangle$) and dash-dotted lines.% The logical error rate per round for the trivalent measurement scheme is consistently slightly larger as for the four-valent measurement scheme.
    The error bars representing the statistical uncertainty are smaller than the markers.}
    \label{fig:s17_comp_schemes}
\end{figure}

\begin{figure}[!t]
    \centering
    \includegraphics[width=\columnwidth]{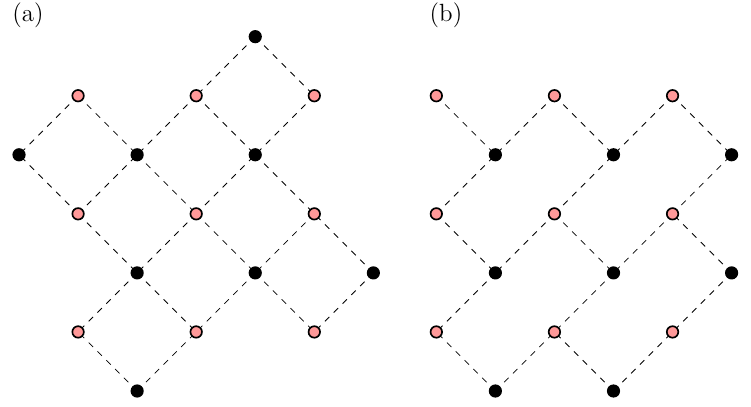}
    \caption{
    Comparison of the qubit layout and connectivity for the distance-$d=3$ surface code in the four-valent (a) and trivalent (b) readout schemes. For distance $d$, the trivalent layout has $(d-1)^2$ physical couplings less compared to the four-valent layout. Data qubits are indicated in red, while ancilla qubits are black. The dashed lines indicate the qubit connectivity that is used to execute the respective scheme.}
    \label{fig:valency_s17}
\end{figure}

As demonstrated in Ref.~\cite{mcewen2023relaxing}, the surface code can operate fault-tolerantly using trivalent circuitry under a standard circuit-level noise model, described in App.~\ref{sec:app_noise}, where all operations are faulty and errors are local. 
Using this model, in which two-qubit gates, measurements, and initializations fail with probability $p$, we confirm numerically for small code distances $d=3,5$,  and $7$, that (i) the trivalent surface code can correct up to $\frac{d-1}{2}$ faults, and (ii) its logical error rate per round is very similar to that of the four-valent circuitry, see~\cref{fig:s17_comp_schemes}.
The corresponding qubit and coupling layout for the $d=3$ rotated surface code is shown in~\cref{fig:valency_s17}. Our results are in qualitative agreement with other works that simulate trivalent QEC under a single-parameter noise model~\cite{mcewen2023relaxing,Benito2025comparativestudyof}.
The memory experiment considered here consists of initializing the data qubits in the state $\ket{0}^{\otimes d^2}$, followed by repeated rounds of stabilizer measurements, and a final measurement in the $Z$ basis. 
The resulting logical error rate $p_L(t)$ after $t$ rounds is used to extract the logical error rate per round $\epsilon_L$~\cite{Brien2017},
\begin{equation}
    p_L=\frac{1}{2}-\frac{1}{2}\left(1-2\epsilon_L\right)^{t+\delta}.
\end{equation}
We estimate $\epsilon_L$ by comparing results for $t=d$ and $t=2d$, yielding $\epsilon_L=(\frac{1-2p_L(2d)}{1-2p_L(d)})^{1/d}$. The parameter $\delta$ accounts for SPAM contributions to the logical error rate.
The circuits use CNOT gates as entangling operations, while measurements and qubit initializations in the $X$ basis replace explicit single-qubit gates and resets. 
An explicit example for $d=3$ is provided in~\cref{fig:memory_circuit}. 
Decoding is performed using \texttt{PyMatching}~\cite{higgott2021}. 
All memory experiments in this section follow this recipe.

\textit{Memory study for realistic noise ---}We now extend this analysis to an experimentally motivated noise model. 
We adopt reference parameters inspired by Ref.~\cite{acharya2024quantum}, representing a superconducting transmon platform with realistic ratios of error rates and operation times. 
Our goal is not to reproduce a specific experiment quantitatively, but to capture the relevant physical structure of the noise.
In addition to gate and measurement errors, we include idling noise arising during gate execution and finite-duration (mid-circuit) measurements~\cite{acharya2024quantum,Moses2023,Graham2023,Krinner2022,wang2026}.
During these periods, qubits accumulate errors that can be modeled by twirled amplitude-damping and dephasing channels~\cite{Tomita2014}. 
This idling noise affects all qubits that are not actively involved in gate operations or measurements during a given time interval. We do not consider additional dynamical decoupling in our circuitry, as its effectiveness depends strongly on the specific implementation and temporal characteristics of the noise~\cite{Viola98,Viola99,Tong2025}.

\begin{figure}[!t]
    \centering
    \includegraphics[width=1\columnwidth]{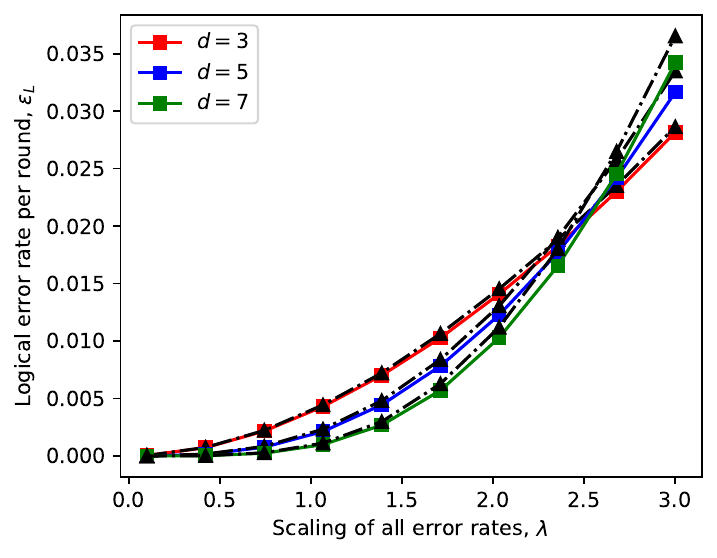}
    \caption{Comparison of the logical error rate per round for the trivalent and the four-valent measurement scheme for an experimentally motivated noise model. A distance $d$ memory experiment for the logical $\ket{0}_L$ state is simulated. 
    The experimentally motivated noise model of~\cref{tab:ref_error_rates} is used where all error parameters and the coherence and life-time are scaled by $\lambda$ as $p\rightarrow\lambda p, T_{1,2}\rightarrow T_{1/2}/\lambda$. The four-valent circuitry is indicated by the colored squares (\cbox{red}) and solid lines while the trivalent variant is given by the corresponding black triangles ($\blacktriangle$) and dash-dotted lines. The logical error rate per round for the trivalent measurement scheme is consistently slightly larger than for the four-valent scheme. The error bars representing the statistical uncertainty are smaller than the markers.}
    \label{fig:experimental_comp_schemes_main}
\end{figure}

Using the reference parameters of~\cref{tab:ref_error_rates}, we find that the logical error rate of the trivalent measurement scheme is slightly increased compared to the four-valent scheme, see~\cref{fig:experimental_comp_schemes_main}. 
To study this systematically, we scale all noise parameters by a factor $0.1 \leq \lambda \leq 1$, such that $T_{1,2} \rightarrow T_{1,2}/\lambda$, which effects the idling noise and $p \rightarrow \lambda p$ for all other error rates. 
The simulations based on this noise model show that the trivalent circuitry exhibits a small structural disadvantage, which we attribute to the changed circuit structure compared to the four-valent circuit and which we analyze further in Appendix \ref{app:noise_susc}.

\textit{Experimental considerations ---} Although a native trivalent hardware implementation is not yet available, the scheme has been demonstrated on a four-valent superconducting platform~\cite{eickbusch2024demonstrating}, showing comparable logical performance. The already competitive performance in experiment and simulations suggests that improved control and mitigation techniques, together with optimized qubit operation in architectures of reduced valency, can lead to improved physical noise levels and therefore to an improved logical performance. %Consequently, one can expect a trivalent hardware implementation to be as competitive as the four-valent one.
In architectures where the qubit capacitance is a limiting resource, a reduced connectivity can enable improved two-qubit gate performance. 
This circumstance is particularly relevant for fluxonium qubits, which require a small self-capacitance to achieve long coherence time. Conversely, it is challenging to devote sufficient capacitance to enable strong capacitive coupling to several neighbors~\cite{Vladimir2009,Rosenfeld2024}.
Reducing the number of connections allows the available capacitance to be redistributed among the three remaining couplings, thereby increasing the coupling strength and enabling faster two-qubit gates~\cite{Ding2023,Blais2021}.
In an idealized scenario, redistributing the capacitance increases the coupling strength by a factor of $4/3$ and reduces the gate time to $3/4$ of its original value. Accordingly, one can expect that the decoherence contribution to the gate infidelity can be reduced by approximately $25\%$.

\begin{figure}[!t]
    \centering
    \includegraphics[width=1\columnwidth]{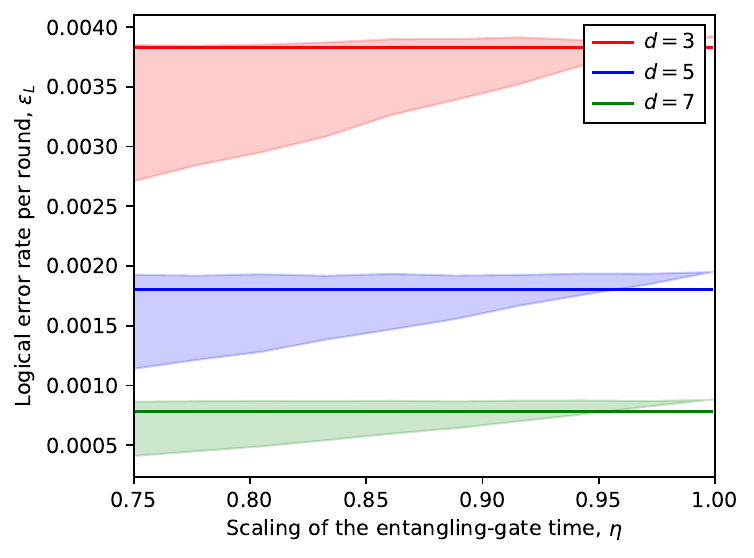}
    \caption{Comparison of the logical error rate for the $\ket{0}_L$ state of the surface code in a memory experiment with the trivalent measurement scheme. The two-qubit gate time is scaled by a factor $\eta$. The gate time is given by $\eta\cdot 47$ns. This also affects the two-qubit error rate as $p_2=(1-\alpha) \cdot 0.4\%+\eta\cdot\alpha\cdot 0.4\%$, which is represented as a cone to cover the range of $\alpha\in[0,1]$. All other error rates are unscaled and correspond to the values given in~\cref{tab:ref_error_rates} for $\lambda=1$. The horizontal dashed lines correspond to the logical error rate of the four-valent measurement scheme for the same noise model at $\eta=1$. The respective colors correspond to different distances as indicated in the legend. }
    \label{fig:gate_time_scaling}
\end{figure}

To model this effect in the numerical memory experiment, we scale the two-qubit gate time $t_2$ by a dimensionless parameter $0.75\leq\eta\leq 1$. We take the reference value $t_2=47\mathrm{ns}$ from superconducting qubit experiments and simulate gate times $t_2=\eta\cdot 47\mathrm{ns}$~\cite{acharya2024quantum}. Following the assumption that this improvement is rooted in the increased interaction strength, we assume that the intrinsic gate quality does not deteriorate.
We model the two-qubit gate error rate by decomposing the probability for a depolarizing error into a base contribution and a contribution that is solely due to decoherence. This decomposition is hard to determine for a concrete experiment. Hence, we tune the decoherence-rooted part of the gate infidelity by a factor $\alpha\in[0,1]$. This part of the gate error rate is consequently reduced once the gate-speed is improved. In summary, we model the two qubit error rate as $p_2=(1-\alpha)0.4\%+\eta\cdot\alpha\cdot 0.4\%$, where $0.4\%$ corresponds to the reference error rate from~\cref{tab:ref_error_rates} and Ref.\cite{acharya2024quantum}. If one considers the bare idling noise during the gate execution as if the qubit was not involved in a gate, one would obtain a contribution value of $\alpha=\mathcal{O}(10^{-1})$. This estimation is expected to underestimate the effect of an improved gate speed. On the other hand, a value of $\alpha\approx1$, will overestimate this effect. Once we scale the gate-speed improvement in~\cref{fig:gate_time_scaling}, we therefore explore the whole range of $\alpha\in[0,1]$.
Furthermore, the gate-time reduction lowers the idling noise of the qubits which are not operated during the entangling gate executions. This effect is covered by the scaling $t_2=\eta\cdot 47\mathrm{ns}$. All other error parameters will not be scaled by $\eta$ and correspond to the previous reference error parameters taken from~\cref{tab:ref_error_rates}.

In~\cref{fig:gate_time_scaling}, we scale $\eta\in[0.75,1]$ and find that for $\eta=1$ the trivalent measurement scheme performs slightly worse than the four-valent one. Assuming that the gate time in the four-valent scheme is not reduced, this disadvantage can be compensated by a gate-time improvement depending on the value of $\alpha$. This range of $\alpha\in[0,1]$ is visualized as a cone. This implies that a larger gate-time improvement puts the trivalent implementation in a better position than the four-valent counterpart with slower gates but otherwise identical noise assumptions.

\begin{table}[!t]
    \centering
    \begin{tabular}{|c|c|}
         \hline
         Two qubit gate error rate& 0.4\% \\
         \hline
         %Single qubit gate error rate& 0.09\%\\
         %\hline
         Measurement/ initialization error rate& 0.8\%\\
         \hline
         \hline
         %Initialization error rate& 0.09\%\\
        %\hline
         Coherence time $T_2$ & 89 $\upmu$s\\
        \hline
         Lifetime $T_1$ & 68 $\upmu$s \\
        \hline
         Two qubit gate time& 47 ns\\
         \hline
         Measurement time for ancilla qubits& 370 ns\\
         \hline
    \end{tabular}
    \caption{Set of physical reference error rates as motivated by the experiment of Ref.~\cite{acharya2024quantum} as the standard for superconducting qubit experiments. Idling errors are modeled by twirled amplitude damping and dephasing channels~\cite{Tomita2014}.}
    \label{tab:ref_error_rates}
\end{table}

\section{Trivalent lattice surgery}
\label{sec:surgery}
Lattice surgery~\cite{Horsman_2012} is a core technique for implementing logical gates in stabilizer codes constrained to planar qubit architectures. 
Lattice surgery enables the creation of logical entanglement and thereby the implementation of logical two-qubit gates, and it also allows for the injection of (non-Clifford) gates based on specific resource states~\cite{litinski2019game}.
Briefly, lattice surgery measures joint logical parities between encoded qubits.
For instance, starting from the state $\ket{0}_L\ket{0}_L$, a joint logical $X_L X_L$ measurement, denoted $M_{XX}$, produces a maximally entangled (Bell) state:
\begin{equation}
\ket{0}_L\ket{0}_L 
\xrightarrow{M_{XX}}
\frac{1}{\sqrt{2}}\Big(\ket{0}_L\ket{0}_L + (-1)^{m_{XX}}\ket{1}_L\ket{1}_L\Big),
\label{eq:bell_state}
\end{equation}
where $m_{XX}=0,1$ is the measurement outcome.

To obtain a deterministic output state, an appropriate logical Pauli correction is applied depending on the value of $m_{XX}$. 
If we replace the first logical qubit with an arbitrary state $\ket{\psi}_L=\alpha\ket{0}_L+\beta\ket{1}_L$, the entanglement generated by the joint measurement can be used to teleport $\ket{\psi}_L$ onto the second qubit by subsequently measuring the logical $Z_L$ parity of the first qubit:
\newline
\begin{align}
\ket{\psi}_L\ket{0}_L 
&\xrightarrow{M_{XX}}
\frac{1}{\sqrt{2}}\Big[(\alpha\ket{0}_L+(-1)^{m_{XX}}\beta\ket{1}_L)\ket{0}_L \nonumber\\
&\quad+(\alpha\ket{1}_L+(-1)^{m_{XX}}\beta\ket{0}_L)\ket{1}_L\Big], \\ \nonumber
&\xrightarrow{M_{ZI}}
(X_L)^{m_{ZI}}\Big(\alpha\ket{0}_L+(-1)^{m_{XX}}\beta\ket{1}_L\Big),
\end{align}
where $m_{ZI}$ is the outcome of the logical $Z_L$ measurement on the first qubit. 
Finally, single-qubit logical Pauli corrections yield the recovered state $\ket{\psi}_L$ on the second qubit. 

The central idea of lattice surgery is to decompose a joint logical Pauli measurement $P_L P_L'$ (e.g., $X_L X_L$ in our example) into measurements of local, low-weight operators. 
For two distance-$d$ codes, the joint logical operator has minimal support on $2d$ physical qubits, making a direct measurement with a single ancilla impractical due to both its non-local extent in the 2D architecture and the risk of correlated error propagation. 

Instead, lattice surgery introduces additional stabilizers that extend the original codes across the boundary between the patches. These stabilizers are defined analogously to those of the surface code and form a mutually commuting set for the enlarged code patch. From this perspective, lattice surgery can be interpreted as \emph{growing} the two code patches into a single, larger code. 

\begin{figure}[!t]
    \centering
    \resizebox{0.8\columnwidth}{!}{%
    \input{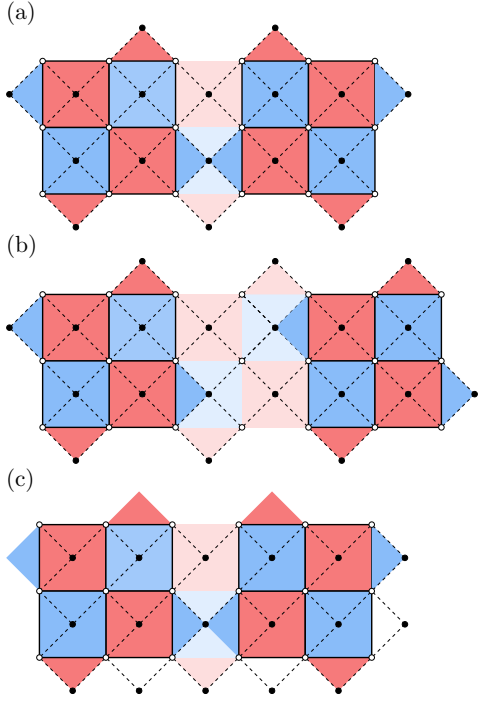}}
    \caption{Comparison of qubit layouts for performing a distance $d=3$ lattice surgery, for which the goal is to measure the joint $XX$ parity (it is given in terms of the product of the light-red plaquette operators). In (a), two four-valent surface code patches are directly next to each other, sharing one $Z$-type boundary ancillary qubit. Upon merging, the $XX$ parity is given as the product of the upper weight four and the lower weight two $X$ stabilizers in the intermediate region. In (b), for the four-valent lattice-surgery scheme, additional data qubits are placed in the intermediate region to extend it and ensure the modularity of the individual patches. The $XX$ parity is given here as the product of the four intermediate $X$-type stabilizers.
    In (c), the trivalent lattice surgery setup is shown with changed qubit connectivity and ancilla arrangement. This layout allows one to operate the two surface code patches in a modular fashion.}
    \label{fig:modularity}
\end{figure}

The desired logical parity $P_L P_L'$ is then obtained as the product of a subset of the measurement outcomes of these newly introduced stabilizers, see~\cref{fig:modularity}. To ensure fault tolerance, the joint parity measurement must be repeated $d$ times
%, requiring at least $\frac{d+1}{2}$ consistent outcomes 
for a distance $d$ code before a logical Pauli correction is assigned. 
Finally, the merged codes can be separated again by remeasuring the original stabilizers of the initial code patches.

In order to design a scalable logical quantum processor, individual logical qubits should be modularly operational~\cite{litinski2019game,boedeker2025}.
For this reason, in the four-valent surface-code design, adjacent logical qubits must be separated by a stripe of $d$ additional data qubits, see~\cref{fig:modularity}(b). 
These $d$ qubits are complemented by $d+1$ ancilla qubits, which are used to merge the two $d\times d$ surface-code patches by projectively initializing a new stripe of surface code of length $d$. 
This constitutes the minimal overhead required to maintain modular operability during lattice surgery. 

The connecting region can be increased in size to enable interactions between code patches that are further apart~\cite{Ramette2024,marton2025latticesurgerybasedlogicalstate}. 
However, it cannot be reduced below this minimal width without sacrificing the independent operability of the two codes.  In particular, in a smaller configuration, the two four-valent patches would have to share a boundary ancilla qubit, as shown in~\cref{fig:modularity}(a).%, thereby coupling their operation.

\begin{figure*}[!t]
    \centering
    \includegraphics[width=2.0\columnwidth]{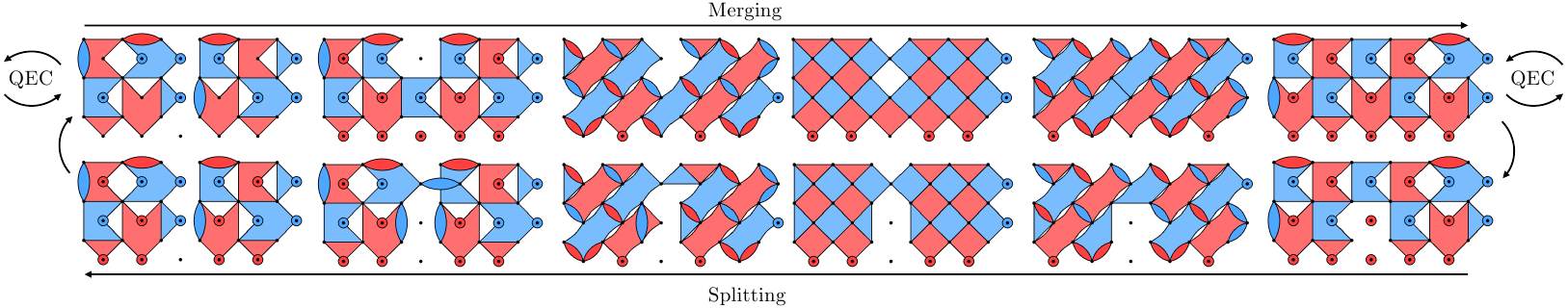}
    \caption{
    Merging and splitting of two surface-17 patches using the trivalent stabilizer-measurement scheme and reduced-overhead qubit layout. 
The figure illustrates how detection regions are deformed during merging, when the stabilizers of the joint code are measured, and how disjoint sets of stabilizers are restored during splitting.}
    \label{fig:triv_merging}
\end{figure*}

In contrast, in the trivalent surface-code design, the boundary ancilla qubits are located on only two sides of each patch, see~\cref{fig:valency_s17}(b). 
As a result, no additional stripe of data qubits is required for the merging operation, while still maintaining modular operability of the two patches. 
The joint parity measurement is obtained from extended boundary stabilizers that effectively fill the region between the patches, see~\cref{fig:triv_merging}. 

Thus, as in the four-valent scheme, merging is performed by executing QEC cycles of the merged code patch, rather than independent QEC cycles of the two separate codes. 
In the trivalent design, the merging region has width of one column of plaquette stabilizers, independent of the code distance.

Specifically, in the trivalent circuit design, every other QEC cycle is conducted with the adjoint circuit of the previous round. 
If two logical surface-code states are initialized and QEC cycles are executed in parallel, then one logical qubit must run the adjoint QEC cycle of the other in order for the two to become compatible for a merging operation. 
As in the four-valent scheme, $\mathcal{O}(d)$ QEC cycles must be performed on the merged code to determine the joint parity value fault-tolerantly. 
Splitting is then achieved by resuming measurement of the stabilizers of the separate code patches. 
As discussed above, this lower-overhead merging operation in the trivalent design implies that fewer ancilla qubits are required compared to the four-valent design. 
The overall difference in overhead between the trivalent and the minimal four-valent merging operation is summarized in~\cref{tab:overhead}. 
Consequently, the trivalent lattice-surgery scheme also requires fewer two-qubit gates during the merging operation. These savings were found independently by Ref.~\cite{low2026denser}.
\begin{comment}
The overall scaling of the number of two-qubit gates with distance is the same for both schemes. However, they differ in the prefactors $\{c_3(d),c_4(d)\}$ governing the leading-order logical error rate,
\begin{equation}
p_L = c_{3,4}(d)\,p^{\frac{d+1}{2}} + \mathcal{O}\!\left(p^{\frac{d+1}{2}+1}\right).
\end{equation}
Here $c_3(d)$ ($c_4(d)$) denotes the prefactor for the trivalent (four-valent) scheme. 
We find a smaller prefactor for the trivalent measurement scheme, i.e., $c_3(d) < c_4(d)$.
Since the two-qubit gate error rate is among the dominant noise sources in state-of-the-art experiments, one can therefore expect the trivalent lattice-surgery protocol to achieve lower logical error rates even under the standard circuit-level noise model.
\end{comment}

\begin{figure}[!t]
    \centering
    \includegraphics[width=\columnwidth]{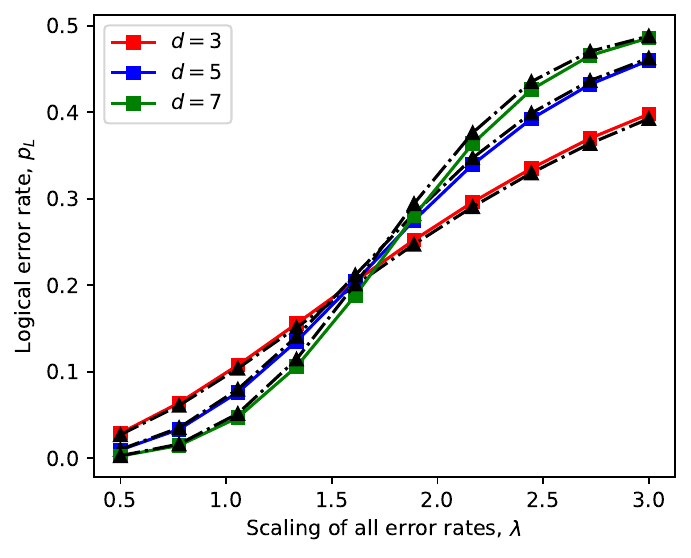}
    \caption{
    Comparison of the logical error rate for teleporting a logical $\ket{0}_L$ state using a trivalent (solid line and $\blacktriangle$) or a four-valent (dash-dotted line and \cbox{red}) lattice surgery protocol for distances $d=3,5,7$. The conducted protocol is summarized in the main text (i-iv). The results are obtained using the experimentally motivated noise model of~\cref{tab:ref_error_rates}, where all error parameters are scaled homogeneously by a parameter $\lambda$. The decoding is performed using MWPM. The error bars representing the statistical uncertainty are smaller than the markers.}
    \label{fig:triv_surg_comp_experimental}
\end{figure}

\begin{figure}[!t]
    \centering
    \includegraphics[width=1\columnwidth]{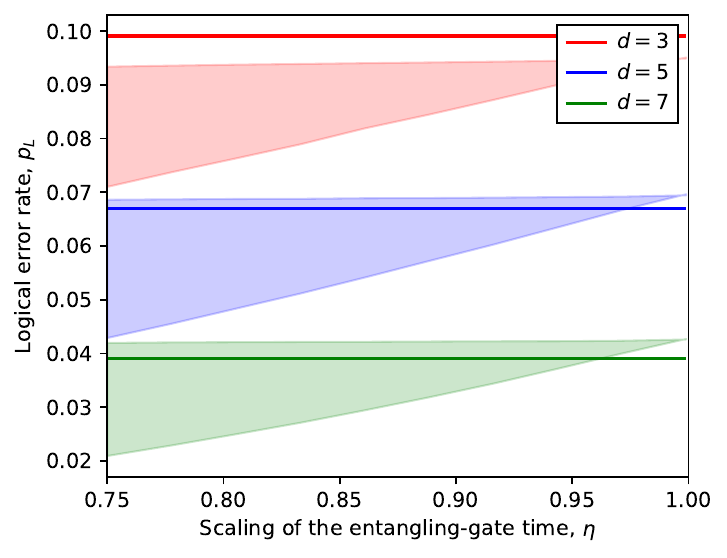}
    \caption{The logical error rate for the teleportation of the $\ket{0}_L$ state of the surface code for the trivalent lattice surgery scheme. The two-qubit gate time is scaled by a factor $\eta$. The gate time is given by $\eta\cdot$$47$ns. This also affects the two-qubit error rate as $p_2=(1-\alpha)0.4\%+\eta\cdot\alpha\cdot 0.4\%$, which is represented as a cone to cover the range of $\alpha\in[0,1]$. All other error rates are unscaled and correspond to~\cref{tab:ref_error_rates} for $\lambda=1$. The horizontal dashed lines correspond to the logical error rate of the four-valent measurement scheme for the same noise model at $\eta=1$. The respective colors correspond to different distances as indicated in the legend. }
    \label{fig:surgery_gate_time_scaling}
\end{figure}

We confirm this conjecture by simulating the creation of a Bell state via a logical $XX$-parity measurement implemented through lattice surgery, see~\cref{eq:bell_state}. 
Additionally, we demonstrate state teleportation using the Bell pair by transferring the initial $\ket{0}_L$ state from the first to the second logical qubit. 
This teleportation protocol consists of:
\begin{enumerate}[(i)]
    \item the projective initialization of the logical state $\ket{0}_L\ket{0}_L$ by performing $d$ QEC cycles, 
    \item the merging of the two code patches by conducting $d$ QEC cycles on the merged code, 
    \item the splitting of the merged code patch by performing a further $d$ QEC cycles on the separate code patches, and 
    \item the transversal measurement of all data qubits in the $Z$ basis. 
    \label{protocol}
\end{enumerate}
\begin{figure}[!t]
    \centering
    \includegraphics[width=\columnwidth]{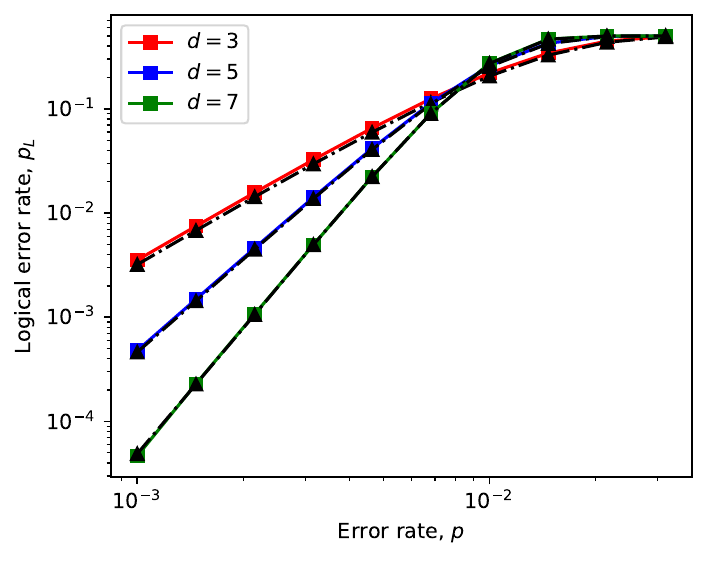}
    \caption{
    Comparison of the logical error rate for teleporting a logical $\ket{0}_L$ state using a trivalent (solid line and $\blacktriangle$) or a four-valent (dash-dotted line and \cbox{red}) lattice surgery protocol for distances $d=3,5,7$. The conducted protocol is summarized in the main text (i-iv). The results are obtained using the standard circuit-level noise model, where errors on two-qubit gates,  measurements, and initializations occur with a probability $p$.  The decoding is performed using MWPM. The error bars representing the statistical uncertainty are smaller than the markers.}
    \label{fig:triv_surg_comp_main}
\end{figure}
The QEC cycles before and after the merging are not strictly necessary to preserve fault tolerance, but they are natural when the protocol is embedded into a larger logical computation. 
We simulate this protocol for both the trivalent and the four-valent schemes using the simple circuit-level noise model, where errors on two-qubit gates,  measurements, and initializations occur with a probability $p$. The results are shown in~\cref{fig:triv_surg_comp_main}. 
We define the logical error probability as the probability that the decoded output has the inverted expected logical $Z_L$ parity of the target logical qubit. This includes a virtual $X_L$ correction based on the measured $Z_L$ parity of the logical source qubit.
Decoding is performed using the minimum-weight perfect matching (MWPM) algorithm implemented in \texttt{PyMatching}~\cite{higgott2021}, with matching weights determined according to the error model. 
The results of~\cref{fig:triv_surg_comp_main} confirm that the logical error rate can be reduced by employing the trivalent circuit design. 
From the low-$p$ behavior, we %extract a stable ratio of $c_{3}(d=3)/c_{4}(d=3)=\textcolor{red}{0.98}$ 
find an improvement of $\approx 2\%$ for distance $d=3$, confirming the reduction of the leading-order prefactor. This ratio increases towards 1 with increasing distance.
For assessing which valency is more suitable for experimental implementations, it is essential to also take into account a realistic noise modeling to judge qualitatively whether an improvement can be expected. We conduct an experimentally motivated noise model study analogous to the one of~\cref{subsec:surf17}, centered around the noise model of~\cite{acharya2024quantum} in~\cref{tab:ref_error_rates}. When scaling all noise parameters by a factor $\lambda$ in~\cref{fig:triv_surg_comp_experimental}, the trivalent lattice surgery performs worse than the four-valent variant.  In particular, only for distance $d=3$, the overhead reduction of the trivalent surgery causes an improvement of the logical teleportation fidelity, but not for distances $d=5,7$ and presumably higher distances. There are two competing effects, on the one side the trivalent circuit performs worse when exposed to the experimentally realistic noise model. On the other side, the trivalent lattice surgery scheme provides a reduction of the number of ancilla qubits and gates that need to be executed for the lattice surgery. This resource saving is, however,  subleading compared to whole protocol when increasing the distance $d$. Further we outline the potential improvement of the trivalent lattice surgery over the four-valent lattice surgery, when a reduced two qubit gate time by a factor $\eta$ is modeled as in the memory case of~\cref{subsec:surf17}. The according result in~\cref{fig:surgery_gate_time_scaling} indicate a clear improvement of the logical error rate of the lattice-surgery-based teleportation of up to $50\%$.

\begin{table}[!t]
    \centering
    \begin{tabular}{|c|c|c|}
    \hline
         & Trivalent& Four-valent \\
         \hline
         Additional data qubits &$0$&$d$\\
         \hline
         Additional ancillary qubits &$1$&$d+1$\\
         \hline
         Number of two-qubit gates& 2$d^2$& $6d^2-2$\\
         \hline
    \end{tabular}
    \caption{Summary of the resources required to perform a merging operation between two distance $d$ surface code patches for trivalent and four-valent operation. The qubit count refers to the minimal number of additional qubits that are needed for the lattice surgery, but not to sustain QEC cycles on the separate patches. The number of two-qubit gates is the count of additional gate executions in the merging region when performing $d$ rounds of QEC cycles of the merged code patch.}
    \label{tab:overhead}
\end{table}

\section{Conclusion}
\label{sec:conclusion}
In this work, we have provided a complementary explanation of the trivalent circuit design for surface-code stabilizer measurements introduced in Ref.~\cite{mcewen2023relaxing}, and confirmed numerically that the trivalent and four-valent measurement schemes exhibit very similar logical performance in the quantum-memory setting under a simple circuit-level noise model that neglects gate durations and idling errors. 
When extending the analysis to a more realistic noise model including idling errors, however, we find that the trivalent measurement scheme exhibits a slightly elevated logical error rate compared to the four-valent scheme. 
In this sense, the trivalent schedule carries a structural disadvantage at the level of repeated QEC cycles when long measurement times and idle periods are taken into account.

At the same time, our results indicate that this disadvantage need not be decisive at the hardware level. 
Reduced valency can improve controllability and operation quality, and may therefore compensate for, or even overcompensate, the additional idling penalty. 
To illustrate this trade-off, we have considered experimentally motivated noise parameters representative of superconducting-qubit platforms~~\cite{acharya2024quantum,Krinner2022,wang2026}. 
Moreover, recent experiments suggest that control-level optimizations can already mitigate the disadvantages associated with the trivalent schedule even on hardware that is not natively trivalent~\cite{eickbusch2024demonstrating}. 
In particular, greater flexibility in arranging qubit and coupler frequencies may facilitate the simultaneous high-quality operation of larger devices.

This perspective is especially relevant for fluxonium-based architectures. 
Because the capacitance budget of each fluxonium qubit must be distributed among its coupling partners, reduced hardware valency is expected to be particularly beneficial~\cite{Ding2023,Blais2021}. 
In such a setting, lower valency may enable both stronger effective couplings and faster two-qubit gates, and thereby improve physical gate performance. 
Our trade-off analysis incorporates this possibility in a simplified manner and indicates that moderate improvements in two-qubit gate speed and quality can already compensate for the idling-related disadvantage of the trivalent measurement scheme.

Beyond repeated QEC cycles, we have shown that the lattice-surgery paradigm developed for the four-valent surface-code layout can be adapted naturally to the trivalent setting. 
Here, the trivalent architecture exhibits a genuine structural advantage: the modified arrangement of boundary ancilla qubits allows lattice surgery to be performed without introducing additional merging data qubits, while preserving modular operability of the logical patches. 
As a result, the merging operation requires fewer auxiliary resources and fewer two-qubit gates than in the minimal four-valent realization. 
This reduces the leading-order prefactor of the logical error rate, and accordingly we have found improved logical performance for logical-state teleportation already under the standard circuit-level noise model.

Taken together, these results show that the advantages of trivalent designs depend on the operational setting under consideration. 
For repeated QEC cycles, trivalent readout can be slightly disadvantaged once realistic idling errors are included. 
For lattice surgery, by contrast, the trivalent layout provides a direct architectural benefit through reduced merging overhead and lower logical error rates. 
Assessing the overall usefulness of trivalent operation therefore requires balancing circuit-level scheduling costs against hardware-level benefits from reduced connectivity.

An important direction for future work is to study trivalent lattice surgery under more hardware-specific noise models and in larger logical processors, where routing and compilation constraints become central. 
More broadly, dynamic circuit designs~\cite{mcewen2023relaxing,eickbusch2024demonstrating}, including schemes in which the support of the surface code is moved between different sets of qubits during QEC cycles, may offer additional opportunities for fault-tolerant compilation and resource optimization.

\section{Acknowledgement}
LB, LC, SG and MM gratefully acknowledge support by the Intelligence Advanced Research Projects Activity (IARPA) and the Army Research Office, under the Entangled Logical Qubits program through Cooperative Agreement Number W911NF-23-2-0212. 
%%%%%%%%%%%%%
LB, LC and MM also acknowledge support by the European Union’s Horizon Europe research and innovation program under Grant Agreement Number 101114305 (``MILLENION-SGA1'' EU Project), the US Army Research Office through Grant Number W911NF-21-1-0007 the Office of the Director of National Intelligence (ODNI), 
and the Deutsche Forschungsgemeinschaft (DFG, German Research Foundation) under Germany’s Excellence Strategy ‘Cluster of Excellence Matter and Light for Quantum Computing (ML4Q) EXC 2004/1’ 390534769.
%%%%%%%%%%%%%
SB acknowledges support from the US National Science Foundation under Award Number OISE-2020174.
%%%%%%%%%%%%%
The authors gratefully acknowledge the computing time provided to
them at the NHR Center NHR4CES at RWTH Aachen
University (project number p0020074).
%%%%%%%%%%%%%
The views and conclusions contained in
this document are those of the authors and should not
be interpreted as representing the official policies, either
expressed or implied, of IARPA, the Army Research Office, or the U.S. Government. The U.S. Government is
authorized to reproduce and distribute reprints for Government purposes notwithstanding any copyright notation
herein.
%%%%%%%%%%%%%
\section{Competing interests}
SG is a founder of Atlantic Quantum Corp., a commercially oriented quantum computing company. The remaining authors declare no competing
interests.

\section{Data availability}
The data and code will be made available upon reasonable request.

\section{Author Contributions}
LB and LC wrote the manuscript with help and feedback from all authors. LB performed the simulations and prepared the figures with input from LC and SB. SG provided insights from the experimental perspective. MM supervised the project.

\appendix
\section{Additional circuits}\label{app:circuit}
We provide example circuits for the results that are presented in the main text.
Generally we have chosen the following design principles:
\begin{itemize}
    \item We use the CNOT gate as the two-qubit gate.
    \item We do not perform resets after ancilla qubit measurements.
    \item We allow for measurements and initializations in the $X$ basis.
    \item No single qubit gates need to be performed.
    \item No echo pulses are inserted.
    \item In the four-valent case we read-out $X$ and $Z$ stabilizers to have a fair comparison to the trivalent measurement scheme.
\end{itemize}
\subsection{Circuits for the memory simulations}
\begin{figure*}[!t]
\centering
\includegraphics[width=\textwidth]{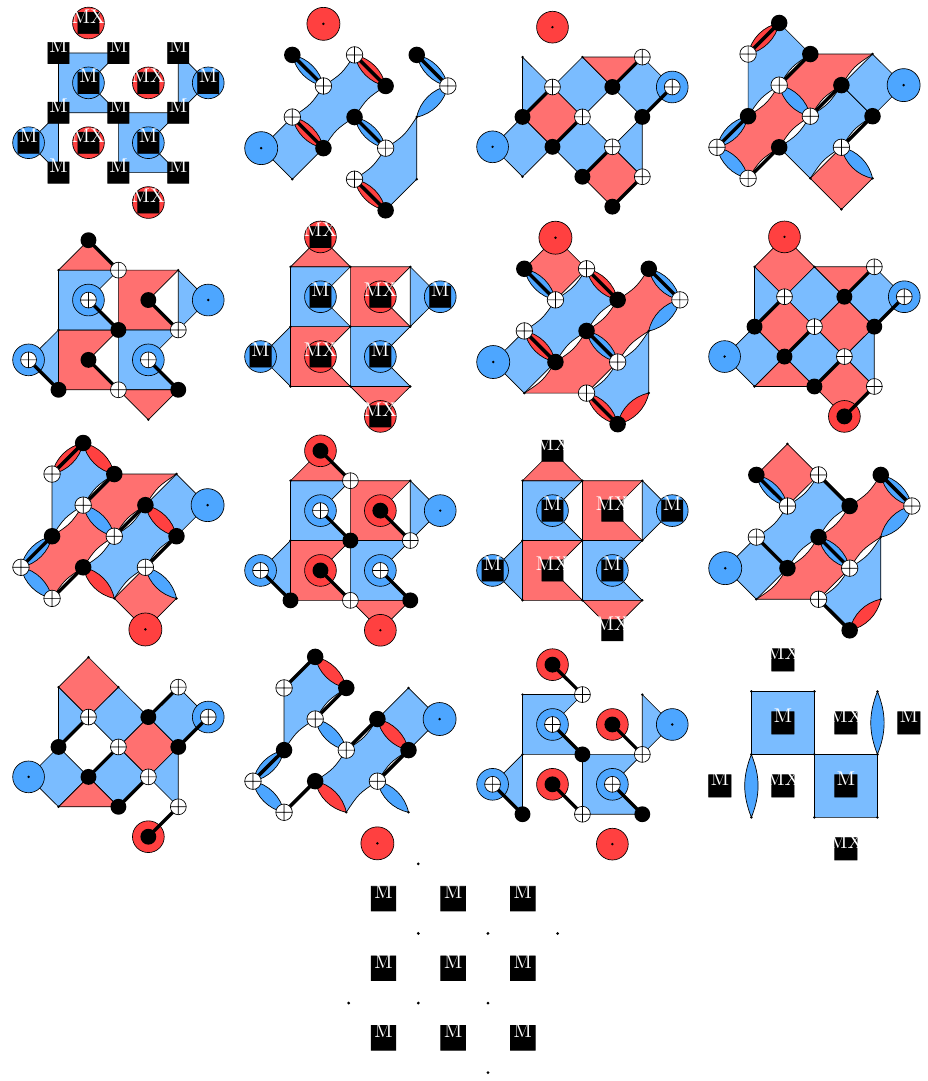}
\caption{Circuitry for conducting three rounds of stabilizer measurements of the $d=3$ rotated surface code using the four-valent circuit design. To be read from left to right and then top to bottom.}
\label{fig:fourv_memory_circuit}
\end{figure*}

In this section we complement the trivalent example circuit for the memory experiments that is discussed in the main text with the respective four-valent circuit. We show the respective distance $d=3$ circuits in~\cref{fig:fourv_memory_circuit}.

\FloatBarrier

\subsection{Circuits for the lattice surgery simulations}

\begin{figure*}[!t]
\centering
\includegraphics[width=0.8\textwidth]{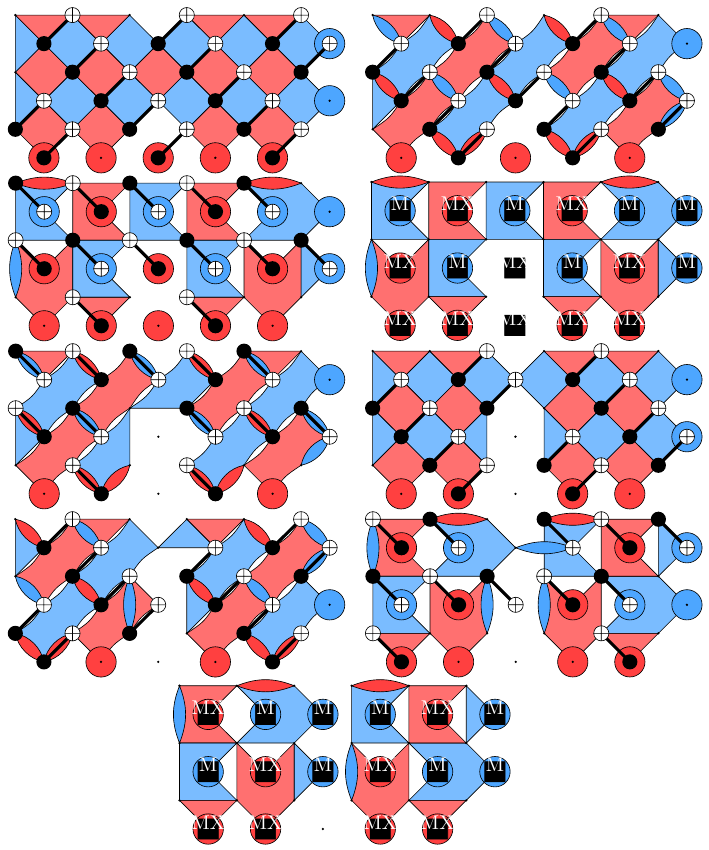}
\caption{Full trivalent surgery circuit that was used for the presented simulations in the main text. To be read from left to right and then top to bottom.}
\label{fig:triv_full_circ}
\end{figure*}

In this section we provide example circuits for the lattice surgery experiments that are discussed in the main text. We show the respective distance $d=3$ circuits in~\cref{fig:triv_full_circ}.

\begin{figure*}[!t]
\centering
\includegraphics[width=0.8\textwidth]{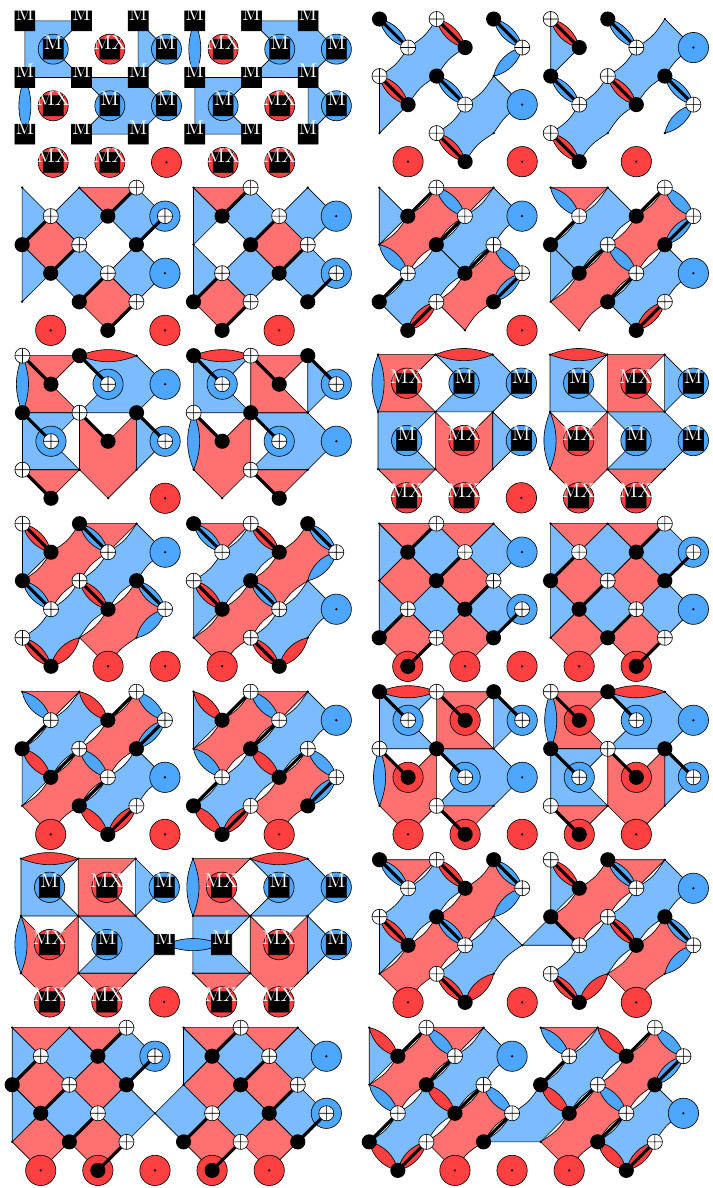}
\end{figure*}

\begin{figure*}[!t]
\centering
\includegraphics[width=0.8\textwidth]{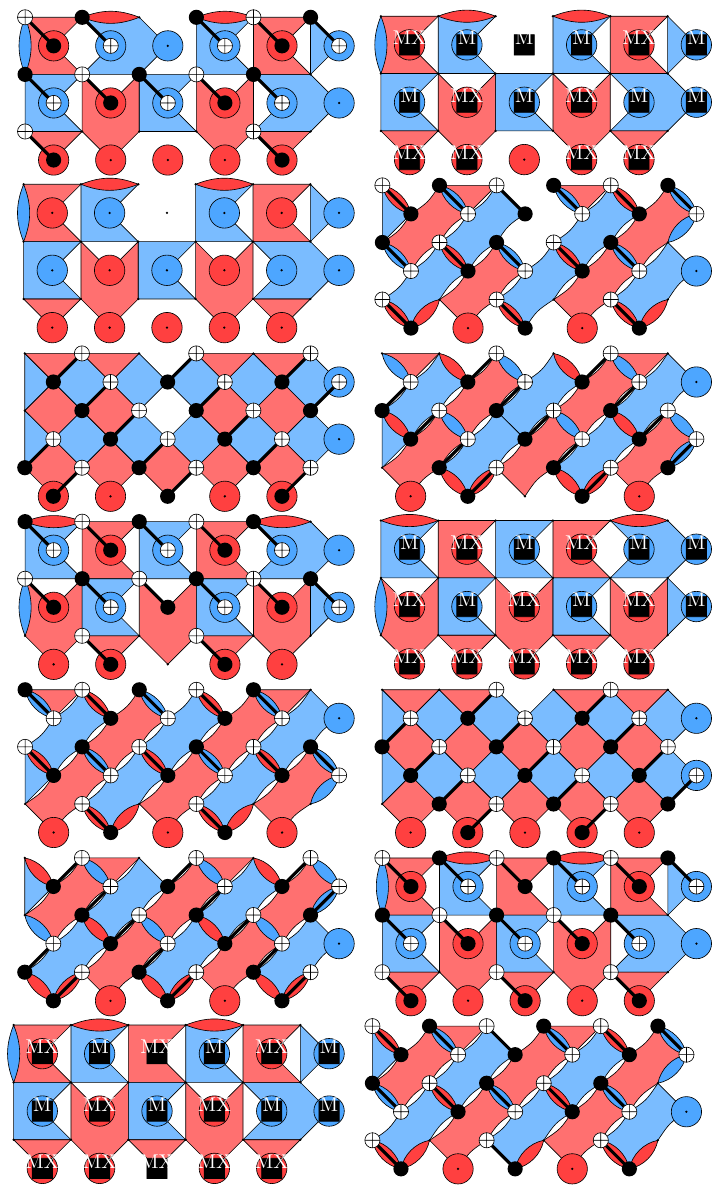}
\end{figure*}

\FloatBarrier
\clearpage
\section{Specifications in the noise models}
\label{sec:app_noise}
We wish to specify in more detail what noise models are used in the simulations of the main text. A major noise source that we consider is the entangling gate infidelity which is modeled by an isotropic two-qubit depolarizing channel with an error probability $p_2$ that is set depending on the simulation setting. The noise channel can be expressed as 
\begin{equation*}
    \rho\rightarrow (1-p_2)\rho +\frac{p_2}{15}\sum_{\substack{P,P'\in\{\mathbb{I},X,Y,Z\} \\ (P,P')\neq\mathbb{I}}} (P\otimes P')\, \rho \,(P\otimes P').
\end{equation*}\\
Measurement errors are modeled as a flip of the qubit state before the measurement with probability $p_{\text{meas}}$. This means that before a measurement in the $Z$ ($X$) basis, there is a probability of $p_{\text{meas}}$ to acquire an $X$ ($Z$) error. We do not employ reset in any circuit.\\
Idling noise is acquired during a time $\tau$; this time corresponds either the two-qubit gate time $t_2$ or the measurement time $t_{\text{meas}}$.
The noise arises from a finite effective coherence time $T_2$ and a finite qubit lifetime $T_1$ that is modeled by a single qubit Pauli channel with error parameters $p_z$, $p_{x/y}$. This channel corresponds to the closest Pauli channel to amplitude damping and dephasing channel~\cite{Tomita2014}
\begin{align*}
    p_z=&
    \frac{1}{2}\left(1-e^{-\tau/T_2}\right)-\frac{1}{4}\left(1-e^{-\tau/T_1}\right)\\
    p_{x/y}=&\frac{1}{4}\left(1-e^{-\tau/T_1}\right).
\end{align*}
Throughout the manuscript we employ three different noise models essentially derived from scaling the three noise components of: two-qubit gate errors, measurement errors and idling errors. We summarize them as
\begin{enumerate}
    \item \textit{Standard noise}: The noise consists of two-qubit gate errors and measurement errors of equal strength $p=p_2=p_{\text{meas}}$. The idling error probability is set to zero $p_z=p_{x/y}=0$. This noise model is used to simulate a memory experiment in~\cref{fig:s17_comp_schemes} and a lattice surgery experiment to perform a logical state teleportation in~\cref{fig:triv_surg_comp_main}.
    \item \textit{Experimentally motivated noise}: The probabilities for two-qubit gate errors and measurement errors as well as idling error probabilities are set according to the values in~\cref{tab:ref_error_rates}. The two-qubit gate error probability and measurement error probability are scaled by the same parameter $\lambda$ as $p_2\rightarrow\lambda p_2$, $p_{\text{meas}}\rightarrow\lambda p_{\text{meas}}$. The idling errors are scaled at the same time by increasing the coherence time and the lifetime of the qubits as $T_2\rightarrow T_2/\lambda$, $T_1\rightarrow T_1/\lambda$. This noise model is used to simulate a memory experiment in~\cref{fig:experimental_comp_schemes_main}.
    \item \textit{Experimentally motivated noise with reduced entangling gate times}: The values for the qubit coherence time and lifetime are taken from~\cref{tab:ref_error_rates} and are left unscaled. The measurement time $t_{\text{meas}}$ is constant as well according to~\cref{tab:ref_error_rates}.
    The measurement error rate is taken as $p_{\text{meas}}=0.8\%$.
    The two-qubit gate time is given by $t_2=\eta\cdot 47$ns, where the parameter $\eta$ is used to model the improvement of the two-qubit gate time $\eta\in [0.75,1]$.
    Correspondingly the idling noise during the execution of two-qubit gates is reduced with decreasing $\eta$. The two-qubit gate error rate is modeled by $p_2=(1-\alpha)0.4\%+\eta\cdot\alpha\cdot 0.4\%$. This shall take into account that the two-qubit gate error is partially due to decoherence during the gate execution and partially due to additional noise. The latter can be imperfect gate calibrations or the coupling to spurious two-level systems. The degree of influence of the gate time for the two-qubit gate error rate is controled by the parameter $\alpha$.
\end{enumerate}

\begin{figure*}[!t]
    \centering
    \includegraphics[width=0.9\textwidth]{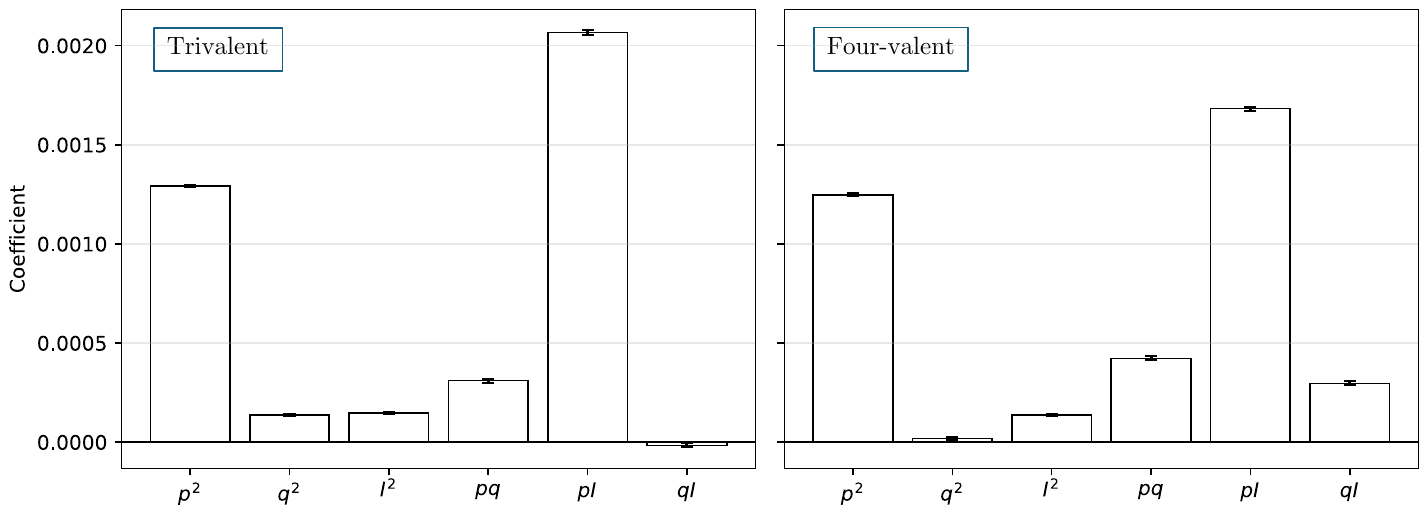}
    \caption{Coefficients of the leading-order quadratic expansion of the logical error rate per round for quantum-memory experiments using the trivalent and four-valent measurement schemes with the reference noise model specified in~\cref{tab:ref_error_rates}. The coefficients are obtained by fitting the expansion in~\cref{eq:expansion} to noise configurations sampled around the reference point. We distinguish three noise sources: the two-qubit gate error rate $p$, the measurement error rate $q$, and the effective idling noise $I$.}
    \label{fig:polynom}
\end{figure*}

\section{Noise susceptibility of the trivalent and four-valent schemes}
\label{app:noise_susc}

We study the influence of different noise sources on the trivalent and four-valent circuit designs using the experimentally realistic noise model specified by the parameters in~\cref{tab:ref_error_rates}. For this purpose, we consider a quantum-memory experiment at distance $d=3$. We focus on three noise sources: the two-qubit gate error rate $p$, the measurement error rate $q$, and the effective idling noise $I$. The idling noise is parametrized by the relaxation and coherence times $T_1$ and $T_2$, respectively.

To vary the strengths of the three noise sources independently, we introduce the scaling parameters $\lambda_p$, $\lambda_q$, and $\lambda_I$. The logical error rate per round is then evaluated as $\epsilon_L(\lambda_p p,\lambda_q q,T_1/\lambda_I,T_2/\lambda_I)$, such that increasing any of the scaling parameters increases the corresponding noise strength. We approximate the dependence of the logical error rate on these scaling parameters by the quadratic expansion
\begin{equation}
\begin{split}
\epsilon_L \approx {}&
c_{p^2}\lambda_p^2
+c_{q^2}\lambda_q^2
+c_{I^2}\lambda_I^2 \\
&+c_{pq}\lambda_p\lambda_q
+c_{pI}\lambda_p\lambda_I
+c_{qI}\lambda_q\lambda_I.
\end{split}
\label{eq:expansion}
\end{equation}

We fit the expansion in~\cref{eq:expansion} using $10\,000$ randomly sampled noise configurations. For each configuration, the scaling parameters are drawn independently from the normal distribution $\lambda_i\sim\mathcal{N}(1,0.1)$, such that the sampled noise configurations lie close to the reference noise model.

The resulting coefficients of the polynomial in~\cref{eq:expansion} are shown in~\cref{fig:polynom} for the trivalent and four-valent measurement schemes. The error bars account for the uncertainty of the fit and the statistical uncertainty of the simulated logical error rates. The coefficients indicate that measurement errors and idling noise affect the trivalent circuit design more strongly than the four-valent circuit design. This difference is also reflected in the mixed contributions that combine measurement or idling errors with two-qubit gate errors.

\printbibliography

\end{document}